\documentclass{aastex61}

\newcommand\aastex{AAS\TeX}

\usepackage{amsmath}

\shorttitle{\aastex\ sample article}
\shortauthors{Tang et al.}

\begin{document}
\title{The properties of Planck Galactic cold clumps in the L1495 dark cloud}

\AuthorCollaborationLimit=200

\author[0000-0001-9160-2944]{Mengyao Tang}
\affil{Department of Astronomy, Yunnan University, and Key Laboratory of Astroparticle Physics of Yunnan Province, Kunming, 650091, China, mengyao\_tang@yeah.net, slqin@bao.ac.cn}

\author{Tie Liu}
\affiliation{Korea Astronomy and Space Science Institute, 776 Daedeokdaero, Yuseong-gu, Daejeon 34055, Republic of Korea, liu@kasi.re.kr}
\affiliation{East Asian Observatory, 660 N. A'ohoku Place, Hilo, HI 96720, USA}

\author{Sheng-Li Qin}
\affiliation{Department of Astronomy, Yunnan University, and Key Laboratory of Astroparticle Physics of Yunnan Province, Kunming, 650091, China, mengyao\_tang@yeah.net, slqin@bao.ac.cn}

\author{Kee-Tae Kim}
\affiliation{Korea Astronomy and Space Science Institute, 776 Daedeokdaero, Yuseong-gu, Daejeon 34055, Republic of Korea, ktkim@kasi.re.kr}

\author{Yuefang Wu}
\affiliation{Department of Astronomy, Peking University, 100871, Beijing China, ywu@pku.edu.cn}

\author{Ken'ichi Tatematsu}
\affiliation{National Astronomical Observatory of Japan, National Institutes of Natural Sciences, 2-21-1 Osawa, Mitaka, Tokyo 181-8588, Japan}

\author{Jinghua Yuan}
\affiliation{National Astronomical Observatories, Chinese Academy of Sciences, Beijing 100101, China}

\author{Ke Wang}
\affil{European Southern Observatory, Karl-Schwarzschild-Str. 2, 85748 Garching bei M\"{u}nchen, Germany}

\author{Harriet Parsons}
\affiliation{East Asian Observatory, 660 N. A'ohoku Place, Hilo, HI 96720, USA}

\author{Patrick M. Koch}
\affiliation{Institute of Astronomy and Astrophysics, Academia Sinica, Taipei, Taiwan}

\author{Patricio Sanhueza}
\affiliation{National Astronomical Observatory of Japan, National Institutes of Natural Sciences, 2-21-1 Osawa, Mitaka, Tokyo 181-8588, Japan}

\author{D. Ward-Thompson}
\affiliation{Jeremiah Horrocks Institute for Mathematics, Physics \& Astronomy, University of Central Lancashire, Preston PR1 2HE, UK}

\author{L. Viktor T\'{o}th}
\affiliation{E\"{o}tv\"{o}s Lor\'{a}nd University, Department of Astronomy, P\'{a}zm\'{a}ny P\'{e}ter s\'{e}t\'{a}ny 1/A, H-1117, Budapest, Hungary}
\affiliation{Konkoly Observatory of the Hungarian Academy of Sciences, H-1121 Budapest, Konkoly Thege Mikl\'{o}s \'{u}t 15-17}

\author{Archana Soam}
\affiliation{Korea Astronomy and Space Science Institute, 776 Daedeokdaero, Yuseong-gu, Daejeon 34055, Republic of Korea}

\author{Chang Won Lee}
\affiliation{Korea Astronomy and Space Science Institute, 776 Daedeokdaero, Yuseong-gu, Daejeon 34055, Republic of Korea}
\affiliation{University of Science \& Technology, 176 Gajeong-dong, Yuseong-gu, Daejeon, Republic of Korea}

\author{David Eden}
\affiliation{Astrophysics Research Institute, Liverpool John Moores University, IC2, Liverpool Science Park, 146 Brownlow Hill, Liverpool L3 5RF, UK}

\author{James Di Francesco}
\affiliation{NRC Herzberg Astronomy and Astrophysics, 5071 West Saanich Rd, Victoria, BC V9E 2E7, Canada}
\affiliation{Department of Physics and Astronomy, University of Victoria, Victoria, BC V8P 5C2, Canada}

\author{Jonathan Rawlings}
\affiliation{Department of Physics and Astronomy, University College London, Gower Street, London, WC1E 6BT, UK}

\author{Mark G. Rawlings}
\affiliation{East Asian Observatory, 660 N. A'ohoku Place, Hilo, HI 96720, USA}

\author{Julien Montillaud}
\affiliation{Institut UTINAM - UMR 6213 - CNRS - Univ Bourgogne Franche Comte, France}

\author{Chuan-Peng Zhang}
\affiliation{National Astronomical Observatories, Chinese Academy of Sciences, Beijing, 100012, China}

\author{M.R. Cunningham}
\affiliation{School of Physics, University of New South Wales, Sydney, NSW 2052, Australia}


\begin{abstract}

Planck Galactic Cold Clumps (PGCCs) possibly  represent the early
stages of star formation. To understand better the properties of PGCCs, we
studied 16 PGCCs in the L1495 cloud with molecular lines and continuum data from \emph{Herschel}, JCMT/SCUBA-2 and the PMO 13.7 m telescope.
Thirty dense cores were identified in 16 PGCCs from 2-D Gaussian fitting. The dense cores have dust
temperatures of $T_{\rm d}$ = 11-14 K, and H$_{2}$ column densities of $N_{\rm H_{2}}$ = 0.36-2.5$\times10^{22}$ cm$^{-2}$.
We found that not all PGCCs contain prestellar objects.
In general, the dense cores in PGCCs are usually at their earliest evolutionary stages.
All the dense cores have non-thermal velocity dispersions
larger than the thermal velocity dispersions
from molecular line data, suggesting that the dense cores may be turbulence-dominated.
We have calculated the virial parameter $\alpha$ and found that 14 of the dense cores have
$\alpha$ $<$ 2, while 16 of the dense cores have $\alpha$ $>$ 2.
This suggests that some of the dense cores are not bound in the absence of external pressure and magnetic fields.
The column density profiles of dense cores were fitted. The sizes of the flat
regions and core radii decrease with the evolution of dense cores.
CO depletion was found to occur in all the dense cores, but is more significant in prestellar core candidates than in protostellar or starless cores.
The protostellar cores inside the PGCCs are still at a very early evolutionary stage, sharing similar physical and chemical properties with the prestellar core candidates.

\end{abstract}




\section{Introduction} \label{sec:intro}
Stars form in dense cores within clumpy and filamentary molecular clouds \citep{An10}.
The dense cores that have no protostars are known as starless cores. When starless cores become dense enough
to be gravitationally bound, they are known as prestellar
cores \citep{Ward-Thompson94}. Then, the prestellar
cores will collapse to form Class 0 and then Class I protostars.
The prestellar cores are compact (with sizes of 0.1~pc or less), cold
($T_{\rm k}\leq$15 K) and dense ($n_{\rm H_{2}}>5\times10^{4}$ cm$^{-3}$) starless
condensations \citep{Caselli11}. \emph{Herschel} observations have revealed that
more than 70\% of the prestellar cores (and protostars) are embedded in
larger, parsec-scale filamentary structures within molecular clouds, which have column
densities exceeding a minimum density threshold ($\sim7\times10^{21}$ cm$^{-2}$) for
core formation \citep{An14}. The properties of prestellar cores, however, are
still not well known due to the lack of a large sample from observations at high
spatial resolution in continuum and molecular lines.
The high frequency channels of \emph{Planck} cover the peak thermal emission
frequencies of dust colder than 14 K. A total of 13188 Planck Galactic
Cold Clumps (PGCCs) were identified \citep{Planck16}. PGCCs have low dust temperatures of 6-20 K, smaller line widths, but modest column densities when compared to other kinds of star forming clouds \citep{Planck11,Planck16, Wu12,Liu13,Liu14}.
A large fraction of PGCCs seem to be quiescent, and not affected by on-going star forming activities \citep{Wu12,Yuan16}.
Those sources are the prime candidates for probing how prestellar cores form and evolve, and for studying the very early stages of star formation across a wide variety of galactic environments.
A large fraction of PGCCs contain YSOs and are forming new stars. As \citet{Zahorecz16} pointed out based on \emph{Herschel} data, about 25\% of PGCC clumps near the Galactic mid-plane may be massive enough to form high-mass stars and star clusters. About 30\% of the Taurus, Auriga, Perseus and California PGCC clumps have associated YSOs \citep{Toth16}. Studying the correlation between the Planck ECC clumps \citep{Planck11} and the AKARI YSOs \citep{Toth14} revealed that 163 of the 915 clumps (17.8\%) have at least one associated YSO within a radius equal to the semi-major axis of the clump.

The properties of the PGCCs are still not well understood due to the relatively poor resolution of the \emph{Planck} telescope.
To understand better the properties of PGCCs, we have been conducting a series of follow-up surveys toward the PGCCs
with several ground-based telescopes such as the PMO (Purple Mountain Observatory) 13.7-m, the TRAO (Taeduk Radio Astronomy Observatory) 14-m, the JCMT (James Clerk Maxwell Telescope) 15-m,
and the NRO (Nobeyama Radio Observatory) 45-m telescopes
\citep{Wu12,Liu12,Meng13,Liu13,Liu14,Liu16,Yuan16,Zhang16,tatematsu17,Kim17,Juvela17}. Using the
SCUBA-2 submillimetre camera on the JCMT 15-m telescope, we have been carrying out a legacy survey
toward ~1000 PGCCs in the 850 $\micron$ continuum, namely ``SCOPE: SCUBA-2 Continuum
Observations of Pre-protostellar Evolution" \citep{Liu17}.
Thousands of dense cores have been identified by the ``SCOPE" survey, and most of them are either starless cores or protostellar cores with very young (Class 0/I) objects.

The L1495 cloud is located at $\sim 140$ pc \citep{Straizys80,Elias87,Kenyon94,Loinard08}, and is representative of a predominantly non-clustered low-mass star formation region.
\citet{Torres12} argued that L1495 is located at a nearer distance of 131.4 pc, but here we retain a distance of 140 pc to be consistent with other studies toward L1495.
The Taurus molecular cloud complex consists of clusters of cold clouds \citep{Toth17}, and dominated by two roughly parallel filamentary structures, as seen for example in the \emph{Herschel} Gould Belt Survey \citep{An10},
L1495 being the most prominent filament of them. A large number
of cold dense cores exist in this region
\citep{Schmalzl10,Ward-Thompson16,Marsh16,Hacar13}. N$_{2}$H$^{+}$
observations toward L1495 performed by \citet{Hacar13} suggest that at least
19 dense cores are embedded in the filamentary cloud.
Other studies such as H$^{13}$CO$^{+}$ performed by \citet{onoshi02}, NH$_{3}$ by \citet{seo15}, and continuum study \citep{Marsh14,Marsh16,Kirk13,Ward-Thompson16} suggested that L1495 cloud is an active low-mass star forming region. The properties of those dense cores in L1495, however, have not been fully investigated before. In this paper, we aim to characterize the properties of a small sample of 16 PGCCs in the L1495 cloud, with data from Herschel, PMO and SCUBA-2. The evolutionary stages, masses, density structures and CO gas depletion of the dense cores inside the PGCCs will be investigated in detail. These studies will deepen our understandings of the physical and chemical properties of those dense cores in L1495 as well as PGCCs in general.

The paper is organized as follows: In Sect. 2, we present the observed data. In
Sect. 3, we describe the observational results. In Sect. 4, we discuss the properties of dense
cores. We summarize our findings and provide a general conclusion in Sect. 5.

\section{Observations and Data}
\subsection{Herschel Data}
The \emph{Herschel} Space Observatory is a 3.5 m-diameter telescope, which operated in the far-infrared and submillimetre regimes \citep{Pilbratt10}. The \emph{Herschel} data of L1495 used in this paper are part of the \emph{Herschel} Gould Belt Survey \citep{An10} and were presented by \citet{Marsh16}.
Details of the observations and data reduction can be seen in
\citet{An10} and \citet{Marsh16}. In this paper, we directly used the column
density and dust temperature maps of L1495 from \citet{Marsh16}, which have a 18$\arcsec$ angular resolution. The column
density maps and dust temperature maps were derived from SPIRE continuum data, by fitting pixel-by-pixel SEDs by using \citep{Hildebrand83}:
\begin{equation}
 F_{\nu} = \frac{N_{\rm H_{2}} m_{\rm H} \mu A B_{\nu}(T)\kappa_{\nu}}{D^{2}},
\end{equation}
where $F_{\nu}$ is the flux density at frequency $\nu$ and B$_{\nu}$
is the Planck Function. $m_{\rm H}$ is the atomic hydrogen mass and $\mu$ is the mean weight of
molecules taken as 2.8 \citep{Kauffmann08}, and \emph{A} is the area of each pixel. A dust mass opacity 0.144 cm$^{2}$ g$^{-1}$ has been derived for $\kappa_{\nu}$ = 0.1$\times$(300/$\lambda_{[\micron]}$)$^{2}$ at 250 $\micron$ and a gas-to-dust mass ratio of 100 are adopted \citep{Hildebrand83,Marsh16}. The pixel size is 6$\arcsec$. The distance \emph{D} is 140 pc \citep{Straizys80,Elias87,Kenyon94,Loinard08}. Therefore, only \emph{T} and $N_{\rm H_{2}}$ are free parameters to fit.
The SED fitting and maps were made by \citet{Marsh16}

\subsection{SCUBA-2 Data}
The Submillimetre Common User Bolometer Array 2 (SCUBA-2) is a bolometer detector operating on the JCMT 15-m telescope with 5120 bolometers in each of two simultaneous imaging bands centred at 450 $\micron$ and 850 $\micron$ \citep{Holland13}.
In this paper we use SCUBA-2 data towards eight of the PGCCs from both the SCOPE survey (M16AL003 and M15BI061; PI: Tie Liu) and CADC \footnote{http://www.cadc-ccda.hia-iha.nrc-cnrc.gc.ca/} archival data. The archival data for G170.26-16.02 (MJLSG37) were actually taken by the SCUBA-2 Gould Belt Legacy Survey \citep{Ward-Thompson07,Buckle15}.
The observations and programs associated with these eight PGCCs are provided in Table 1. The data were observed primarily using the CV Daisy mode. The CV Daisy is designed for small compact sources providing a deep 3$\arcmin$ region in the centre of the map but coverage out to beyond 12$\arcmin$ \citep{Bintley14}.
We used CV Daisy mode in the SCOPE survey because this mode is more efficient to quickly survey a large sample. The aim of SCOPE survey is to detect dense condensations inside PGCCs.
All the SCUBA-2 850 $\micron$ continuum data were reduced using an iterative map-making technique \citep{Chapin13,Currie14,Mairs15}. Specifically the data were all run with the same reduction tailored for compact sources, filtering out scales larger than 200$\arcsec$ on a 4$\arcsec$ pixel scale. A Flux Conversion Factor (FCF) of 554 Jy/pW/beam was used to convert data from pW to Jy/beam \citep{Liu17}.
The FCF in this paper is higher than the canonical value derived by \citet{Dempsey13}.
This higher value reflects the impact of the data reduction technique and pixel size used in by the authors.
The pixel size used in the reduction of a calibrator can have a significant effect on the FCF derived. The effect is different for both the beam and aperture FCFs, and also for different calibrators\citep{Dempsey13}. Therefore, we derived a new FCF for the SCOPE survey for a default pixel size of 4$\arcsec$. We should note that the flux calibration uncertainty in SCOPE survey is less than 10\% at 850 $\micron$ band. The archival data for G170.26-16.02 and G171.91-15.65 were calibrated with a FCF of 537 Jy/pW/beam. The FCF used for those archival data are consistent with the values (528 Jy/pW/beam for G170.26-16.02 and 526 Jy/pW/beam for G171.91-15.65) derived from the calibrators observed at the same time. We note that G170.26-16.02 was observed with Pong1800 mode. The filtering out scale for G170.26-16.02 data reduction is 600 $\arcsec$. However, different filtering out scale would not affect the core properties because the core sizes in G170.26-16.02 are much smaller than 100$\arcsec$.

\begin{deluxetable}{ccccc}[h!] 
\setlength{\tabcolsep}{0.05in}
\tabletypesize{\scriptsize}  \tablecaption{The SCUBA-2 observations}
 \tablewidth{0pt} \tablehead{
  Name & project ID & UT &obs number(s)  &rms\tablenotemark{a} (mJy/beam)
}\startdata
G168.13-16.39                  &M16AL003                  &2015-12-17   &33            &11.93 \\
G168.72-15.48                  &M16AL003                  &2016-01-13   &13            &7.18 \\
G169.76-16.15\tablenotemark{b} &M16AL003                  &2015-12-27   &36            &12.37 \\
G170.00-16.14\tablenotemark{b} &M16AL003                  &2015-12-27   &36            &13.03 \\
G170.26-16.02                  &MJLSG37\tablenotemark{c}  &2014-11-16   &21,25 $\&$ 24 &12.50 \\
G171.49-14.90                  &M16AL003                  &2016-01-17   &20            &9.24 \\
G171.80-15.32                  &M15BI061\tablenotemark{d} &2015-09-28   &15            &9.22 \\
G171.91-15.65                  &JCMTCAL\tablenotemark{e}  &2012-02-11   &33            &8.25 \\
\enddata
\tablenotetext{a}{The rms in the final reduced map.}
\tablenotetext{b}{PGCCs G169.76-16.15 and G170.00-16.14 are covered within single observation.}
\tablenotetext{c}{These observations were PONG 1800 observations observed as part of the JCMT Gould Belt Survey \citep{Ward-Thompson07,Buckle15}.}
\tablenotetext{d}{M15BI061 is a pilot study of the SCOPE survey.}
\tablenotetext{e}{JCMT calibration observation of object DGTau.}
\end{deluxetable}

\subsection{PMO 13.7 m Telescope Data}
Observations of the 16 PGCCs in L1495 in the $^{12}$CO(1-0),$^{13}$CO(1-0), and
C$^{18}$O(1-0) lines were performed with the PMO 13.7 m telescope between January and May of 2011.
The 9-beam array receiver system in double-sideband (DSB) mode was used as the front end \citep{Shan12}.
The $^{12}$CO(1-0) line was observed in the upper sideband and both $^{13}$CO(1-0) and C$^{18}$O(1-0) were observed simultaneously in the lower sideband.
The half-power beam width was 56$\arcsec$ with a main beam efficiency of 50$\%$. The pointing and tracking accuracies were both better than 5$\arcsec$.
The spectral resolution was $\sim$61 KHz, corresponding to a velocity resolution of 0.16 km s$^{-1}$.

The on-the-fly (OTF) observing mode was used. OTF data were converted to three-dimensional datacubes with a grid spacing of 30$\arcsec$. Then, we used MIRIAD \citep{Sault95} for further analyses. A description of the analyses of the OTF data can be found in \citet{Liu12} and \citet{Meng13}

\section{RESULTS}
The distributions of the 16 PGCCs are shown in Figure 1, which clearly shows the filamentary structure in L1495. The names and coordinates of the 16 PGCCs are listed in Table 2.

\begin{figure} 
\plotone{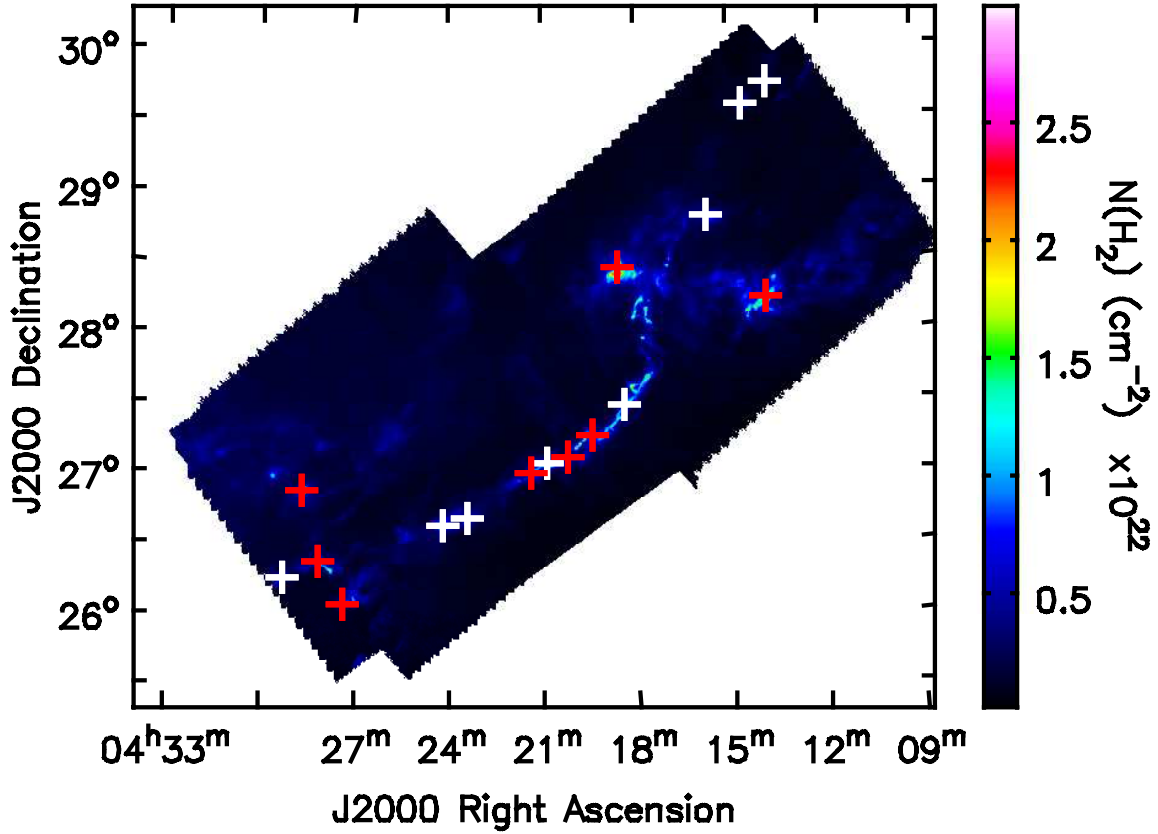}
\caption{Spatial distribution of the observed 16 PGCCs which are marked with crosses. The red cross represents the PGCC with SCUBA-2 detection, while the white cross represents the PGCC without SCUBA-2 detection. The colour image is the \emph{Herschel} H$_{2}$ column density map, which was derived from \emph{Herschel} continuum data by fitting SEDs pixel-by-pixel \citep{Marsh16}. An obvious filamentary structure in L1495 is visible.}
\end{figure}
\begin{deluxetable}{cccccccc}[h!] 
\setlength{\tabcolsep}{0.05in}
\tabletypesize{\scriptsize}  \tablecaption{The coordinates of observed Planck cold clumps}
 \tablewidth{0pt} \tablehead{
  Name& Glon & Glat &Ra(J2000)  &Dec(J2000) &$N_{\rm VC}$\tablenotemark{a} &$N_{\rm core}$\tablenotemark{b} &$N_{\rm condensation}$\tablenotemark{c} \\
      &($\arcdeg$) &($\arcdeg$) &(h m s) &(d m s) }
\startdata
G166.99-15.34     & 166.99217    & -15.346058       & 04 13 42.01  & +29 44 26.31  &2 &1 &0 \\
G167.23-15.32     & 167.23387    & -15.326718       & 04 14 30.92  & +29 35 16.09  &1 &1 &0 \\
G168.00-15.69     & 168.00291    & -15.694481       & 04 15 40.46  & +28 48 00.55  &1 &1 &0 \\
G168.13-16.39     & 168.13475    & -16.393131       & 04 13 47.56  & +28 13 22.16  &2 &2 &3 \\
G168.72-15.48     & 168.72801    & -15.481486       & 04 18 34.58  & +28 26 34.64  &1 &2 &3 \\
G169.43-16.17     & 169.43114    & -16.179394       & 04 18 23.08  & +27 28 18.97  &2 &2 &0 \\
G169.76-16.15     & 169.76073    & -16.159975       & 04 19 25.47  & +27 15 16.21  &2 &5 &3 \\
G170.00-16.14     & 170.00243    & -16.140558       & 04 20 12.04  & +27 05 52.89  &2 &2 &2 \\
G170.13-16.06     & 170.13426    & -16.062908       & 04 20 50.61  & +27 03 29.08  &2 &1 &0 \\
G170.26-16.02     & 170.2661     & -16.024094       & 04 21 21.47  & +26 59 29.26  &1 &2 &3 \\
G170.83-15.90     & 170.83739    & -15.907699       & 04 23 24.42  & +26 39 55.87  &2 &2 &0 \\
G170.99-15.81     & 170.9912     & -15.810754       & 04 24 10.41  & +26 37 16.68  &1 &1 &0 \\
G171.49-14.90     & 171.49657    & -14.901693       & 04 28 39.71  & +26 51 56.81  &2 &1 &2 \\
G171.80-15.32     & 171.80418    & -15.326718       & 04 28 07.26  & +26 21 41.65  &1 &3 &3 \\
G171.91-15.65     & 171.91405    & -15.655738       & 04 27 20.24  & +26 03 50.27  &1 &3 &2 \\
G172.06-15.21     & 172.06786    & -15.210718       & 04 29 15.63  & +26 14 52.01  &1 &1 &0 \\
\enddata
\tablenotetext{a}{$N_{\rm VC}$ represents the number of velocity components.}
\tablenotetext{b}{$N_{\rm core}$ represents the number of dense cores which detected by \emph{Herschel}.}
\tablenotetext{c}{$N_{\rm condensation}$ represents the number of dense condensations which detected by SCUBA-2.}
\end{deluxetable}

\subsection{Identification and Classification of Dense Cores}
We identified dense cores within the observed PGCCs based on the column density ($N_{\rm H_{2}}$) maps derived from \emph{Herschel} data by eye.
We did not apply any core finder algorithm because the emission peaks in the Taurus PGCCs can be easily identified by eye.
Then the identified dense cores were fitted with 2-D Gaussian to get the physical parameters (e.g., size, temperature, density).
We mainly focus on the dense cores with core-averaged column densities larger than 3$\times$10$^{21}$ cm$^{-2}$. The density threshold
we used is about half of the minimum density threshold ($\sim$7$\times$10$^{21}$ cm$^{-2}$) for core formation discovered in \emph{Herschel} observations\citep{An14}. Their peak column densities are also larger than $\sim$7$\times$10$^{21}$ cm$^{-2}$.
In total 30 most reliable dense cores with mean column densities larger than 3$\times$10$^{21}$ cm$^{-2}$ are identified in 16 clumps from \emph{Herschel} data.

As mentioned in section 1, dense cores that have no protostars are classified as starless cores. When starless cores become dense enough to be gravitationally bound, they become prestellar cores \citep{Ward-Thompson94}. Then, the dense cores will collapse to form Class 0 and then Class I protostars.

To investigate how core properties change with evolution, we classify the dense cores into starless, prestellar, and protostellar categories.
The presence of a \emph{Herschel} 70 $\micron$ source is a signpost of ongoing star formation \citep{Konyves15}. Therefore, those dense cores with 70 $\micron$ emission are protostellar. In contrast, those having no 70 $\micron$ emission are starless. As an example, Figure 2 presents the \emph{Herschel} column density map (in contours) overlaid on \emph{Herschel} 70 $\micron$ emission map for a representative source PGCC G171.91-15.65. The cores ``H1" and ``H2" in G171.91-15.65 are associated with protostars, while ``H3" is starless. In this paper, we only show images for G171.91-15.65. The images for other PGCCs are shown in Appendix.

Prestellar cores are gravitationally bound starless cores and show higher density than unbound starless cores.
Many factors like gravity, turbulence, magnetic field, external pressure, and even bulk motions can affect the stability of dense cores. Therefore, it is hard to tell whether or not a starless core is gravitationally bound based on present line data. The dense starless cores detected by SCUBA-2, however, having higher density than other starless cores (e.g. those without SCUBA-2 detection), should be good candidates for prestellar cores \citep{Ward-Thompson16}.
Therefore, in this paper, we classify the starless cores with SCUBA-2 detection as prestellar core candidates.
In total, we identify 9 protostellar cores, 6 prestellar core candidates and 15 starless cores.
The starless cores, prestellar core candidates, and protostellar cores are marked with 1, 2, or 3 in column 10 of Table 3, respectively.

\begin{figure} 
\plotone{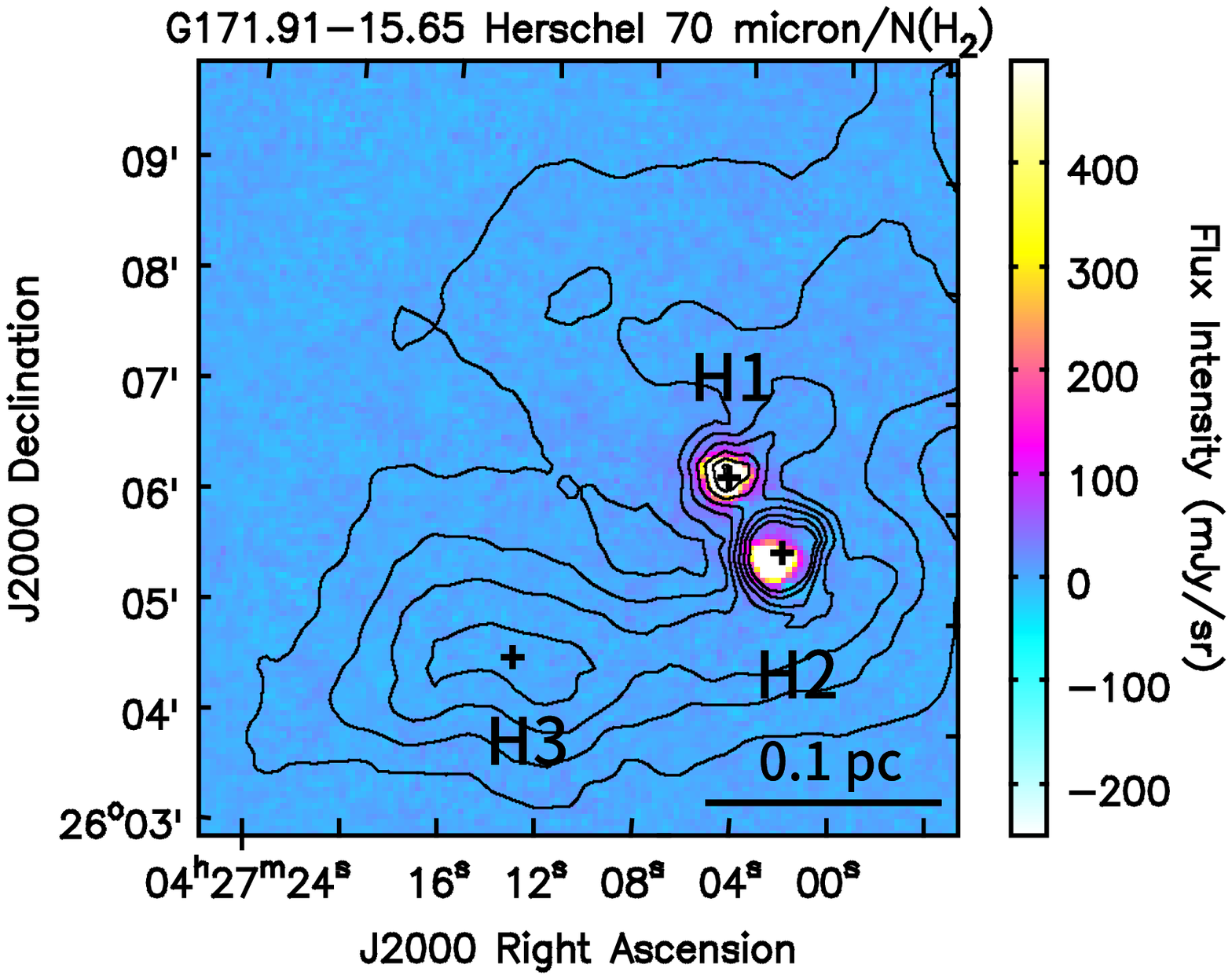}
\caption{The \emph{Herschel} column density map has overlaid on \emph{Herschel} 70 $\micron$ continuum emission map for the representative source PGCC G171.91-15.65. The images for other sources are shown in Appendix.
Contours represent the column density distribution from \emph{Herschel}.
The contour levels are from 30\% to 90\% of peak value, in step 10\%. The \emph{Herschel} 70 $\micron$ continuum are shown in colour.
Protostellar cores ``H1" and ``H2" are associated with 70 $\micron$ sources, while starless core ``H3" is not associated with any 70 $\micron$ source.}
\label{}
\end{figure}

\subsection{Herschel Images}
The \emph{Herschel} column density distributions and dust temperature maps are shown in contours and color in panel a of Figure 3, respectively.
The coordinates and FWHM deconvolved major and minor axes of the dense cores (a and b) were obtained by 2-D Gaussian fits.
The effective radius is \emph{R} =$\sqrt{ab}D$, where \emph{D} is the distance.
The core-averaged $N_{\rm H_2}$(\emph{Herschel}) and $T_{\rm d}$ of dense cores are derived within 2 FWHM (or 2\emph{R}) area, and presented in Table 3.
The core-averaged $N_{\rm H_2}$(\emph{Herschel}) of the dense cores range from 3.6($\pm$1.0)$\times$10$^{21}$ to 2.5($\pm$0.6)$\times$10$^{22}$ cm$^{-2}$. The core-averaged dust temperatures of the dense cores range from 11.1($\pm$0.4) to 13.8($\pm$0.4) K.
If we assume that the dense core is a sphere with a radius of $R$, the volume density of the core is roughly $n_{\rm H_2} = N_{\rm H_{2}}^{\rm peak}/2R$, where $N_{\rm H_{2}}^{\rm peak}$ is peak column density of dense core.
The volume densities ($n_{\rm H_2}$) of the dense cores range from 5.7($\pm$0.9)$\times$10$^{3}$ cm$^{-3}$ to 7.9($\pm$1.4)$\times$10$^{4}$ cm$^{-3}$.
The masses of the dense cores are estimated as:
\begin{equation}
\it{M_{\rm core} = \frac{\rm 4}{\rm 3}\pi R^{\rm 3}\cdot n_{\rm H_{2}}\cdot m_{\rm H}\cdot \mu},
\end{equation}
where $\mu$ = 2.8 is the mean molecular weight \citep{Kauffmann08}. $m_H$ is the mass of a H atom.
The core masses $M_{\rm core}$ are presented in Table 3.

\begin{figure} 
\plotone{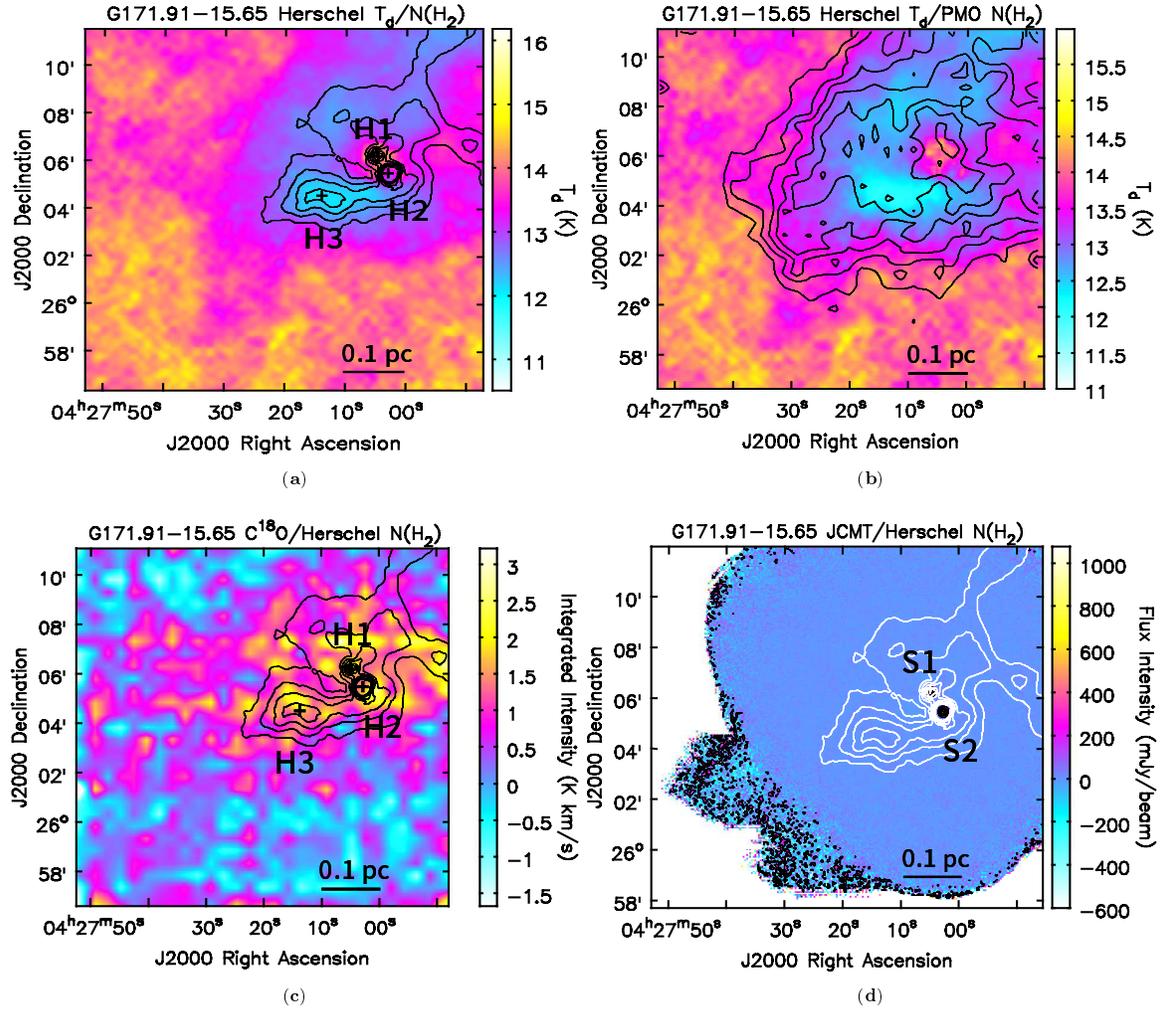}
\caption{\emph{Panel a}: The \emph{Herschel} column density map have overlaid on the dust temperature map for the representative source PGCC G171.91-15.65. Contours represent the column density distributions from \emph{Herschel} data. The contour levels are from 30\% to 90\% of peak value, in step 10\%.
\emph{Panel b}: The PMO H$_{2}$ column density map (derived from $^{13}$CO data) have overlaid onto the dust temperature map for the representative source PGCC G171.91-15.65. Contours represent the column density distribution from PMO data. The contour levels are from 30\% to 90\% of peak value, in step 10\%. The dust temperatures are shown in colour in K.
\emph{Panel c}: The \emph{Herschel} column density map have overlaid onto the C$^{18}$O integrated intensity map for the representative source PGCC G171.91-15.65.
The contours represent the column density distributions from \emph{Herschel} data.
The contour levels are from 30\% to 90\% of peak value, in step 10\%. The integrated intensities of C$^{18}$O data are shown in colour in K km s$^{-1}$.
\emph{Panel d}: The \emph{Herschel} column density map have overlaid on SCUBA-2 850 $\micron$ continuum emission map for the representative source PGCC G171.91-15.65. White contours represent the \emph{Herschel} column density distributions. The contour levels are from 30\% to 90\% of peak value, in step 10\%. The SCUBA-2 850 $\micron$ continuum intensities are shown in colour and black contours. The contour levels are from 10 \% to 90 \% of peak value, in step 20 \%. The condensations detected by SCUBA-2 are marked with -S1, -S2 or -S3. Only two condensations are detected by SCUBA-2 in PGCC G171.91-15.65 and their sizes are smaller than the dense cores detected by \emph{Herschel}. The same maps for other sources are shown in Appendix.}
\label{}
\end{figure}

\begin{deluxetable}{ccccccccccccccccccccccccccccccccccccccc} 
\tabletypesize{\tiny} \tablecolumns{18}
\tablewidth{0pc} \setlength{\tabcolsep}{0.05in}
\tablecaption{The parameters derived from Herschel data} \tablehead{
 Name & \multicolumn{2}{c}{Coordinate(J2000)}  &Deconvolved Size &P.A\tablenotemark{a} &$N_{\rm H_{2}}$(Herschel) &$N_{\rm H_{2}}^{\rm peak}$(Herschel) &$T_{d}$  & \emph{R}\tablenotemark{b} & $n_{\rm H_{2}}$ &$M_{\rm core}$  &Classification\tablenotemark{c} & Remark\tablenotemark{d} \\
 & \multicolumn{2}{c}{(Ra,Dec)}    &("$\times$") &($\degr$)    &(10$^{22}$ cm$^{-2}$)  &(10$^{22}$ cm$^{-2}$) &(K) &(pc)  &(10$^{4}$ cm$^{-3}$) &(M$_{\sun}$) & &
}\startdata
G166.99-15.34-H1 &\multicolumn{2}{c}{(4:13:41.951,+29:43:43.811)} &214$\times$176 &-65 &0.36($\pm$0.10) &0.62 &13.1($\pm$0.4) &0.11     &0.90($\pm$0.14) &3.6($\pm$0.6)  &1 & \\
G167.23-15.32-H1 &\multicolumn{2}{c}{(4:14:29.120,+29:35:14.313)} &233$\times$112 &78  &0.75($\pm$0.24) &1.19 &12.5($\pm$0.4) &0.11     &1.75($\pm$0.35) &6.7($\pm$1.3)  &1 & \\
G168.00-15.69-H1 &\multicolumn{2}{c}{(4:15:35.466,+28:47:23.381)} &480$\times$307 &-86 &0.59($\pm$0.14) &0.92 &12.7($\pm$0.3) &0.26     &0.57($\pm$0.09) &29.2($\pm$4.4) &1 & \\
G168.13-16.39-H1 &\multicolumn{2}{c}{(4:13:49.122,+28:12:31.360)} &382$\times$164 &-40 &1.53($\pm$0.39) &2.54 &11.9($\pm$0.3) &0.17     &2.42($\pm$0.37) &34.6($\pm$5.3) &3 &S1 \\
G168.13-16.39-H2 &\multicolumn{2}{c}{(4:14:08.510,+28:09:24.391)} &627$\times$164 &-40 &1.33($\pm$0.14) &1.64 &11.6($\pm$0.2) &0.22     &1.22($\pm$0.11) &36.4($\pm$3.1) &3 & \\
G168.72-15.48-H1 &\multicolumn{2}{c}{(4:18:33.568,+28:26:56.805)} &394$\times$158 &-6  &1.40($\pm$0.22) &1.97 &11.9($\pm$0.3) &0.17     &1.89($\pm$0.21) &26.5($\pm$3.0) &1 & \\
G168.72-15.48-H2 &\multicolumn{2}{c}{(4:18:38.298,+28:23:22.959)} &251$\times$169 &-85 &2.46($\pm$0.60) &4.13 &13.2($\pm$1.0) &0.14     &4.78($\pm$0.70) &37.8($\pm$5.5) &2 &S1 \\
G169.43-16.17-H1 &\multicolumn{2}{c}{(4:18:20.859,+27:28:27.755)} &616$\times$152 &-43 &1.05($\pm$0.23) &1.49 &11.9($\pm$0.3) &0.21     &1.16($\pm$0.18) &30.3($\pm$4.7) &1 & \\
G169.43-16.17-H2 &\multicolumn{2}{c}{(4:18:36.843,+27:22:20.914)} &309$\times$137 &-8  &0.99($\pm$0.18) &1.35 &11.8($\pm$0.3) &0.14     &1.57($\pm$0.21) &12.3($\pm$1.7) &1 & \\
G169.76-16.15-H1 &\multicolumn{2}{c}{(4:19:01.227,+27:16:50.760)} &215$\times$119 &-27 &0.97($\pm$0.12) &1.17 &11.9($\pm$0.2) &0.11     &1.74($\pm$0.18) &6.5($\pm$0.7)  &1 & \\
G169.76-16.15-H2 &\multicolumn{2}{c}{(4:19:12.104,+27:13:53.438)} &232$\times$129 &-37 &0.95($\pm$0.09) &1.09 &12.2($\pm$0.1) &0.12     &1.51($\pm$0.12) &7.1($\pm$0.6)  &1 & \\
G169.76-16.15-H3 &\multicolumn{2}{c}{(4:19:24.655,+27:15:05.860)} &206$\times$124 &-48 &1.05($\pm$0.26) &1.70 &12.1($\pm$0.3) &0.11     &2.53($\pm$0.39) &9.4($\pm$1.5)  &2 &S3 \& 21 \\
G169.76-16.15-H4 &\multicolumn{2}{c}{(4:19:37.299,+27:15:19.802)} &170$\times$144 &-69 &1.11($\pm$0.24) &1.66 &11.8($\pm$0.3) &0.11     &2.53($\pm$0.37) &8.8($\pm$1.3)  &1 &22 \\
G169.76-16.15-H5 &\multicolumn{2}{c}{(4:19:42.843,+27:13:32.947)} &142$\times$78  &-75 &1.13($\pm$0.35) &1.84 &12.0($\pm$0.3) &0.072    &4.17($\pm$0.79) &4.4($\pm$0.8)  &3 &S1 \& 23 \\
G170.00-16.14-H1 &\multicolumn{2}{c}{(4:19:51.580,+27:11:33.739)} &108$\times$75  &-35 &1.24($\pm$0.30) &1.82 &11.1($\pm$0.3) &0.062    &4.82($\pm$0.78) &3.2($\pm$0.5)  &2 &S2 \& 24 \\
G170.00-16.14-H2 &\multicolumn{2}{c}{(4:19:58.119,+27:10:16.257)} &144$\times$79  &-44 &1.25($\pm$0.42) &2.32 &12.2($\pm$0.8) &0.065    &5.82($\pm$1.05) &4.6($\pm$0.8)  &3 &S1 \& 25 \\
G170.13-16.06-H1 &\multicolumn{2}{c}{(4:20:54.832,+27:02:42.492)} &400$\times$188 &-89 &1.16($\pm$0.26) &1.64 &11.5($\pm$0.3) &0.19     &1.43($\pm$0.22) &26.8($\pm$4.3) &1 &30, 31 \& 32 \\
G170.26-16.02-H1 &\multicolumn{2}{c}{(4:21:12.207,+27:01:17.035)} &101$\times$50  &-51 &0.99($\pm$0.17) &1.22 &12.7($\pm$0.4) &0.049    &4.08($\pm$0.56) &1.3($\pm$0.2)  &3 &S2 \& 34 \\
G170.26-16.02-H2 &\multicolumn{2}{c}{(4:21:21.509,+26:59:53.620)} &192$\times$85  &-52 &1.24($\pm$0.38) &2.19 &11.1($\pm$0.4) &0.087    &4.08($\pm$0.71) &7.8($\pm$1.4)  &2 &S3 \& 33 \\
G170.83-15.90-H1 &\multicolumn{2}{c}{(4:23:37.832,+26:40:21.545)} &204$\times$162 &48  &0.53($\pm$0.13) &0.70 &12.2($\pm$0.3) &0.12     &0.93($\pm$0.16) &5.0($\pm$0.9)  &1 & \\
G170.83-15.90-H2 &\multicolumn{2}{c}{(4:23:29.382,+26:38:35.041)} &180$\times$110 &50  &0.54($\pm$0.08) &0.94 &12.2($\pm$0.3) &0.096    &1.58($\pm$0.13) &4.0($\pm$0.4)  &1 & \\
G170.99-15.81-H1 &\multicolumn{2}{c}{(4:24:17.134,+26:36:54.550)} &669$\times$308 &-75 &0.82($\pm$0.18) &1.35 &11.8($\pm$0.3) &0.31     &0.71($\pm$0.10) &59.8($\pm$8.1) &1 &35 \& 36 \\
G171.49-14.90-H1 &\multicolumn{2}{c}{(4:28:39.189,+26:51:43.701)} &135$\times$92  &-25 &2.12($\pm$0.67) &3.70 &11.3($\pm$0.4) &0.076    &7.87($\pm$1.43) &10.0($\pm$1.8) &3 &S1 \\
G171.80-15.32-H1 &\multicolumn{2}{c}{(4:28:09.290,+26:20:44.816)} &148$\times$97  &-60 &1.50($\pm$0.51) &2.52 &11.1($\pm$0.4) &0.082    &5.01($\pm$1.01) &7.9($\pm$1.6)  &2 &S1 \& 39 \\
G171.80-15.32-H2 &\multicolumn{2}{c}{(4:27:54.940,+26:19:17.994)} &97$\times$82   &70  &0.89($\pm$0.15) &1.19 &12.4($\pm$0.7) &0.061    &3.17($\pm$0.40) &2.0($\pm$0.3)  &3 &S2 \& 38 \\
G171.80-15.32-H3 &\multicolumn{2}{c}{(4:27:48.456,+26:18:07.157)} &237$\times$122 &59  &1.34($\pm$0.25) &1.81 &11.1($\pm$0.2) &0.12     &2.54($\pm$0.32) &11.3($\pm$1.6) &2 &S3 \& 37 \\
G171.91-15.65-H1 &\multicolumn{2}{c}{(4:27:04.261,+26:06:22.493)} &86$\times$48   &36  &0.78($\pm$0.26) &1.35 &13.8($\pm$0.4) &0.044    &4.99($\pm$0.94) &1.2($\pm$0.2)  &3 &S1 \\
G171.91-15.65-H2 &\multicolumn{2}{c}{(4:27:02.366,+26:05:24.059)} &90$\times$61   &30  &0.98($\pm$0.39) &2.07 &13.3($\pm$0.6) &0.051    &6.63($\pm$1.26) &2.5($\pm$0.5)  &3 &S2 \\
G171.91-15.65-H3 &\multicolumn{2}{c}{(4:27:13.117,+26:04:30.520)} &161$\times$125 &-72 &0.82($\pm$0.09) &0.98 &12.3($\pm$0.1) &0.096    &1.64($\pm$0.16) &4.2($\pm$0.4)  &1 & \\
G172.06-15.21-H1 &\multicolumn{2}{c}{(4:29:15.723,+26:14:00.580)} &215$\times$113 &-72 &0.87($\pm$0.26) &1.47 &12.1($\pm$0.3) &0.11     &2.24($\pm$0.40) &7.7($\pm$1.4)  &1 & \\
\enddata
\tablenotetext{a}{Position angle of dense cores which detected by \emph{Herschel}, and the convention used for measuring angles is east of north.}
\tablenotetext{b}{The core radius are derived from elliptic Gaussian fitting to \emph{Herschel} dense cores}
\tablenotetext{c}{Classification of dense cores, 1 represents starless core, 2 represents prestellar candidate, 3 represents protostellar core.}
\tablenotetext{d}{S1, S2 or S3 represent the SCUBA-2 counterpart of dense cores. The numbers is source ID of NH$_{\rm 3}$ counterpart from \citet{seo15}.}
\end{deluxetable}

\subsection{ SCUBA-2 Continuum Images}
Panel d of Figure 3 presents the 850 $\micron$ map from SCUBA-2 overlaid on \emph{Herschel}
column density map. The SCUBA-2 detected condensations are marked with -S1, -S2 or -S3.
The SCUBA-2 observations filtered out the large scale ($>200\arcsec$) extended emission and only picked up the dense condensations inside the cores.
Some dense cores show very flatten and extended structure, and thus their emissions are mostly filtered out in SCUBA-2 observations.
In total, we identified 22 condensations with signal-to-noise ratios larger than 3 from SCUBA-2 images.
According to panel d of Figure 3,  the SCUBA-2 detected condensations are consistently smaller than that detected by \emph{Herschel}.
The total integrated fluxes of these condensations were calculated from 2-D Gaussian fits. The masses were consequently obtained using \citet{Kauffmann08}:
\begin{eqnarray}
M = &0.12~{\rm M_{\sun}}\left(e^{1.439(\lambda/ \rm mm)^{-1}(\it T/ \rm 10~K)^{-1}}-1\right) \nonumber \\
&\left(\frac{\kappa_{\nu}}{0.01 \rm ~cm^{2} g^{-1}}\right)^{-1}\left(\frac{S_{\nu}}{\rm Jy}\right)\left(\frac{D}{100 \rm ~pc}\right)^{2}\left(\frac{\lambda}{\rm mm}\right)^{3}
\end{eqnarray}
where \emph{T} is temperature, adopting the
\emph{Herschel} dust temperature in our calculation, \emph{D} is the distance to the source,
$\kappa_{\nu}$=0.012 is the dust opacity at 850 $\micron$ wavelength, which is consistent with the value used in SED fit to the \emph{Herschel} data in section 2.1.
\emph{S}$_{\nu}$ is the total flux density of the core region. The masses of the 22 SCUBA-2 condensations range from
0.02($\pm$0.01) M$_{\sun}$ to 1.25($\pm$0.13) M$_{\sun}$ and the column
densities range from 4.2($\pm$0.5)$\times$10$^{21}$ cm$^{-2}$ to 8.1($\pm$0.3)$\times$10$^{22}$
cm$^{-2}$. The volume densities range from 1.5($\pm$0.2)$\times$10$^{4}$ cm$^{-3}$ to
1.4($\pm$0.06)$\times$10$^{6}$ cm$^{-3}$. All core parameters derived from SCUBA-2 data are presented in Table 4.

SCUBA-2 can detect the condensations denser than that detected by \emph{Herschel}.
The \emph{Herschel} volume densities of dense cores range from
5.7($\pm$0.9)$\times$10$^{3}$ to 7.9($\pm$1.4)$\times$10$^{4}$ cm$^{-3}$.
The volume densities derived from SCUBA-2 data range from 7.4($\pm$1.5)$\times$10$^{3}$ to
9.1($\pm$0.6)$\times$10$^{5}$ cm$^{-3}$. The volume density estimated from the SCUBA-2 data is larger than the value estimated from the \emph{Herschel} data, indicating that the SCUBA-2 detected condensations are denser than dense cores detected by \emph{Herschel}.
Figure 4 presents the correlation between $M_{\rm condensation}$(SCUBA-2) and $M_{\rm core}$(\emph{Herschel}).
The gray area represent mass ratios which range from 0.025 to 0.125.
The green squares and red diamonds represent prestellar candidates and protostellar cores, respectively.
All presetllar candidates in Figure 4 are covered by gray area, but most protostellar cores show mass ratios higher than 0.125.
This indicates that protostellar cores are generally more concentrated than prestellar candidates.
This may indicate that as core evolve, the volume density distributions of dense cores will be more centrally concentrated.

\citet{Ward-Thompson16} identified 25 dense condensations in L1495 cloud based on SCUBA-2 observations.
According to their results, the mass of dense condensations ranges from 0.02 to 0.61 M$_{\sun}$, with a mean value of 0.19.
Their radii ranges from 0.02 to 0.03 pc, with a mean value of 0.03 pc.
It should be noted that they have taken $\beta$ as 1.3 to derived $\kappa_{\nu}$ in their calculations. If $\beta$=2 was taken, their condensations' masses should be doubled and are very similar to our SCUBA-2 detected condensations in PGCCs (mass ranges from 0.02 to 1.25 M$_{\sun}$, with a mean value of 0.45 M$_{\sun}$, radii ranges from 0.02 to 0.06 pc, with a mean value of 0.03 pc.)

\begin{figure} 
\plotone{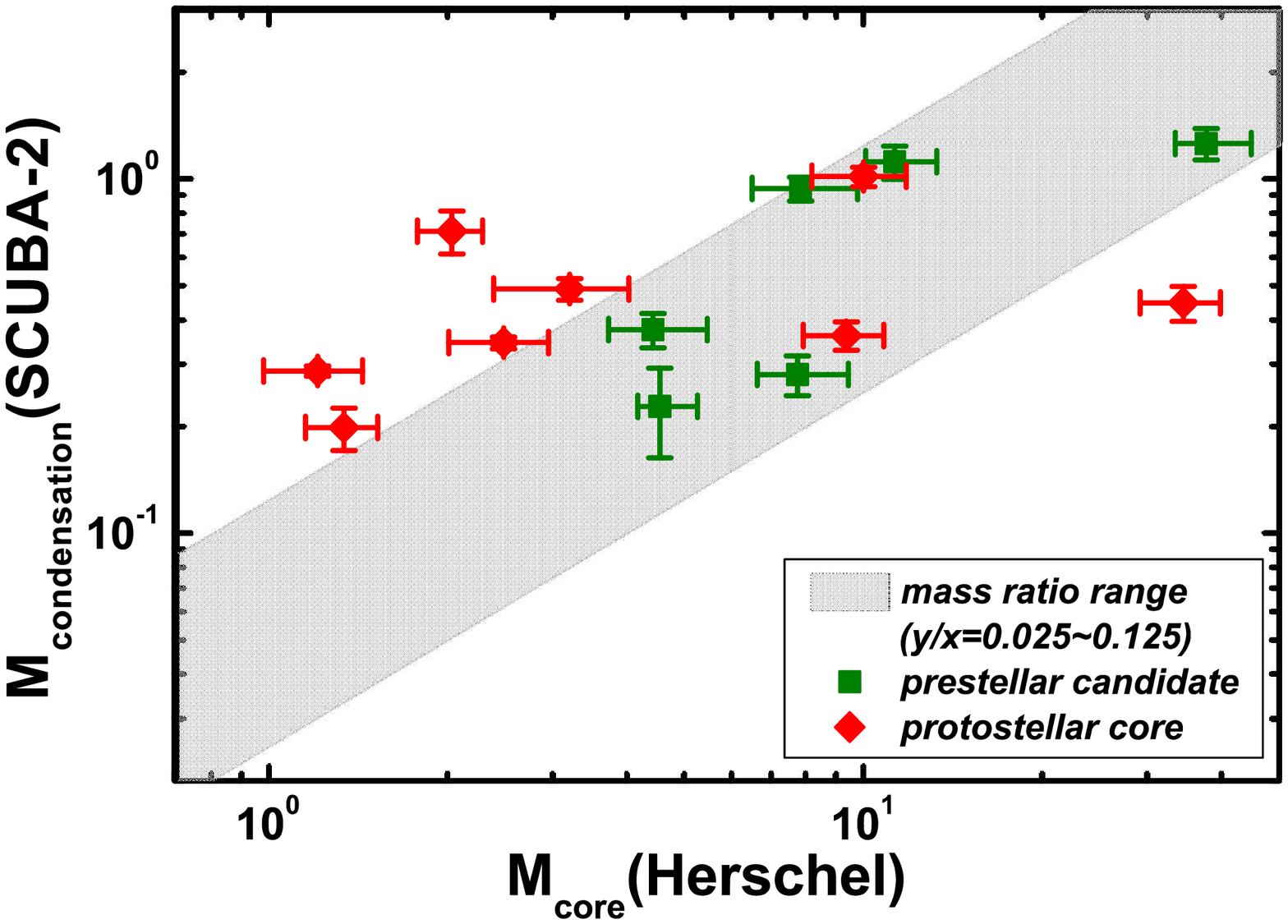}
\caption{The correlation between $M_{\rm core}$(\emph{Herschel}) and $M_{\rm condensation}$(SCUBA-2). The green squares and red diamonds represent prestellar candidates and protostellar cores, respectively. The gray area is mass ratios range (0.025$\sim$0.125). All prestellar candidates are covered by gray area, but most protostellar cores show mass ratio higher than 0.125.}
\label{}
\end{figure}

\begin{deluxetable}{cccccccccccccccccccccccccccccccccccc} 
\tabletypesize{\tiny} \tablecolumns{12}
\tablewidth{0pc} \setlength{\tabcolsep}{0.05in}
\tablecaption{The parameters of the dense cores derived from SCUBA-2 850 $\micron$ data} \tablehead{
 Name & \multicolumn{2}{c}{Coordinate(J2000)}   &\colhead{Total flux} &Peak flux density &Deconvolved size &P.A\tablenotemark{a} & \emph{R}(SCUBA-2)\tablenotemark{b} &\colhead{$N_{\rm H_{2}}$(SCUBA-2) } & $n_{\rm H_{2}}$  & \emph{$M_{\rm condensation}$}  &Remark\tablenotemark{c} \\
 & \multicolumn{2}{c}{(Ra,Dec)}  &(mJy)  &(mJy) &("$\times$") &($\degr$) &(10$^{-2}$ pc) &(10$^{21}$ cm$^{-2}$)  &(10$^{4}$ cm$^{-3}$) &\colhead{(M$_{\sun}$)}
}\startdata
G168.13-16.39-S1 &\multicolumn{2}{c}{(4:13:48.350,+28:12:31.114)}   &1096($\pm$140)  &80.40  &72$\times$50  &20  &4.10  &5.63($\pm$0.65)  &2.23($\pm$0.26)   &0.44($\pm$0.05)     &H1     \\
G168.13-16.39-S2 &\multicolumn{2}{c}{(4:13:54.792,+28:11:33.400)}   &66($\pm$20)     &71.60  &14$\times$14  &0   &0.95  &5.39($\pm$1.14)  &9.20($\pm$1.95)   &0.02($\pm$0.01)     &      \\
G168.72-15.48-S1 &\multicolumn{2}{c}{(4:18:39.688,+28:23:24.070)}   &4038($\pm$182)  &183.43 &89$\times$51  &-79 &4.63  &12.49($\pm$1.27) &4.39($\pm$0.45)   &1.25($\pm$0.13)     &H2     \\
G168.72-15.48-S2 &\multicolumn{2}{c}{(4:18:39.116,+28:21:56.082)}   &1229($\pm$139)  &75.57  &116$\times$37 &-77 &4.50  &4.16($\pm$0.44)  &1.50($\pm$0.16)   &0.40($\pm$0.04)     &      \\
G168.72-15.48-S3 &\multicolumn{2}{c}{(4:18:03.409,+28:22:52.354)}   &1424($\pm$195)  &102.70 &126$\times$35 &84  &4.55  &4.74($\pm$0.59)  &1.69($\pm$0.21)   &0.46($\pm$0.06)     &      \\
G169.76-16.15-S1 &\multicolumn{2}{c}{(4:19:42.596,+27:13:38.231)}   &1166($\pm$128)  &424.45 &23$\times$20  &-74 &1.48  &34.88($\pm$3.22) &38.20($\pm$3.53)  &0.40($\pm$0.03)     &H5     \\
G169.76-16.15-S2 &\multicolumn{2}{c}{(4:19:41.612,+27:16:08.184)}   &177($\pm$37)    &144.54 &14$\times$14  &0   &0.95  &12.50($\pm$1.83) &21.36($\pm$3.14)  &0.05($\pm$0.01)     &      \\
G169.76-16.15-S3 &\multicolumn{2}{c}{(4:19:23.793,+27:14:54.189)}   &1034($\pm$125)  &76.65  &84$\times$37  &-74 &3.83  &5.42($\pm$0.60)  &2.29($\pm$0.25)   &0.37($\pm$0.04)     &H3     \\
G170.00-16.14-S1 &\multicolumn{2}{c}{(4:19:58.676,+27:09:58.420)}   &1346$\pm$174)   &396.04 &26$\times$24  &58  &1.74  &34.05($\pm$4.53) &31.69($\pm$4.22)  &0.48($\pm$0.06)     &H2     \\
G170.00-16.14-S2 &\multicolumn{2}{c}{(4:19:51.105,+27:11:27.498)}   &632($\pm$110)   &81.79  &59$\times$41  &58  &3.36  &4.29($\pm$0.65)  &2.07($\pm$0.31)   &0.23($\pm$0.03)     &H1     \\
G170.26-16.02-S1 &\multicolumn{2}{c}{(4:21:07.522,+27:02:31.323)}   &257($\pm$67)    &56.54  &31$\times$25  &-49 &1.93  &5.01($\pm$1.09)  &4.22($\pm$0.92)   &0.09($\pm$0.02)     &      \\
G170.26-16.02-S2 &\multicolumn{2}{c}{(4:21:11.866,+27:01:13.907)}   &578($\pm$81)    &83.59  &49$\times$21  &-54 &2.19  &8.74($\pm$1.20)  &6.47($\pm$0.88)   &0.20($\pm$0.03)     &H1     \\
G170.26-16.02-S3 &\multicolumn{2}{c}{(4:21:21.236,+26:59:53.873)}   &798($\pm$96)    &70.25  &70$\times$25  &-54 &2.91  &7.03($\pm$0.90)  &3.92($\pm$0.50)   &0.28($\pm$0.04)     &H2     \\
G171.49-14.90-S1 &\multicolumn{2}{c}{(4:28:39.417,+26:51:37.576)}   &2734($\pm$150)  &249.51 &60$\times$45  &-13 &3.56  &17.00($\pm$1.08) &7.75($\pm$0.49)   &1.01($\pm$0.06)     &H1     \\
G171.49-14.90-S2 &\multicolumn{2}{c}{(4:29:04.907,+26:49:08.135)}   &200($\pm$29)    &170.36 &14$\times$14  &0   &0.95  &14.21($\pm$1.53) &24.28($\pm$2.61)  &0.06($\pm$0.01)     &      \\
G171.80-15.32-S1 &\multicolumn{2}{c}{(4:28:10.207,+26:20:28.341)}   &2667($\pm$155)  &186.27 &91$\times$61  &63  &5.07  &7.73($\pm$0.60)  &2.47($\pm$0.19)   &0.93($\pm$0.07)     &H1     \\
G171.80-15.32-S2 &\multicolumn{2}{c}{(4:27:57.598,+26:19:25.435)}   &1735($\pm$169)  &146.73 &67$\times$46  &64  &3.81  &10.43($\pm$1.46) &4.45($\pm$0.62)   &0.71($\pm$0.10)     &H2     \\
G171.80-15.32-S3 &\multicolumn{2}{c}{(4:27:49.314,+26:18:09.353)}   &2871($\pm$155)  &65.58  &136$\times$52 &63  &5.76  &7.15($\pm$0.78)  &2.02($\pm$0.22)   &1.11($\pm$0.12)     &H3     \\
G171.91-15.65-S1 &\multicolumn{2}{c}{(4:27:04.776,+26:06:18.335)}   &858($\pm$21)    &88.37  &14$\times$14  &0   &0.95  &67.36($\pm$2.33) &115.07($\pm$3.97) &0.29($\pm$0.01)     &H1     \\
G171.91-15.65-S2 &\multicolumn{2}{c}{(4:27:02.654,+26:05:32.941)}   &1092($\pm$8.33) &108.65 &14$\times$14  &0   &0.95  &80.95($\pm$3.23) &138.30($\pm$5.53) &0.34($\pm$0.01)     &H2     \\
\enddata
\tablenotetext{a}{Position angle of condensations which detected by SCUBA-2, and the convention used for measuring angles is east of north.}
\tablenotetext{b}{The radius are derived from elliptic Gaussian fitting to SCUBA-2 condensations}
\tablenotetext{c}{H1, H2, H3 or H5 represent the \emph{Herschel} counterpart of condensations.}
\end{deluxetable}

\subsection{ Results of PMO Molecular Lines}

The $^{12}$CO(1-0), $^{13}$CO(1-0), and C$^{18}$O(1-0) lines were simultaneously observed
with the PMO telescope.
The core-average $^{12}$CO(1-0), $^{13}$CO(1-0), and C$^{18}$O(1-0) spectra of 30 \emph{Herschel} dense cores are presented in Figure 5.
The spectra of $^{12}$CO(1-0), $^{13}$CO(1-0), and  C$^{18}$O(1-0) are shown in red, green and blue, respectively.
From Gaussian fitting, we obtain peak velocity, full width of half maximum (FWHM) and peak brightness temperature.
Some dense cores (17 dense cores) show multiple velocity components and for these we fitted multiple Gaussian components.
The fitting parameters of all dense cores are presented in Table 5.
For line with multiple velocity components, the velocity component is determined to be associated with the \emph{Herschel} dense core if its integrated intensity map shows similar morphology to its \emph{Herschel} map.

The excitation temperature $T_{\rm ex}$ of $^{12}$CO is calculated following \citep{Garden91}:
\begin{eqnarray}
 T_{\rm b} =& \frac{T_{\rm a}^{*}}{\eta_{b}}=\frac{h\nu}{k}\left[\frac{1}{e^{(h\nu/k{T_{\rm ex}})}-1}-\frac{1}{e^{(h\nu/k{T_{\rm bg}})}-1}\right]\times \nonumber \\
&\left(1-e^{-\tau}\right)f,
\end{eqnarray}
where $T_{\rm b}$ is the brightness temperature. $T_{\rm bg}$ is the background temperature of 2.73 K. $T_{\rm a}^{*}$ is the observed antenna temperature, \emph{k} is the Boltzmann constant, $\eta_{\rm b}$=0.6 is the main-beam efficiency, and $\nu$ is the frequency. Assuming that $^{12}$CO(1-0) is optically thick ($\tau \gg 1$) and that the filling factor \emph{f} = 1, then $T_{\rm ex}$ can be obtained. The mean, maximum, minimum, median $T_{\rm ex}$ values of each dense core are presented in Table 6. The mean $T_{\rm ex}$ ranges from 8.5 K to 15.1 K, consistent with the dust temperature range.

If we assume that $^{12}$CO(1-0) and $^{13}$CO(1-0) have the same excitation temperature, the $^{13}$CO optical depth can be obtained. These values are also presented in Table 6. Then, the $^{13}$CO column density can be calculated as follows \citep{Garden91}:
\begin{equation}
\it{N = \frac{{\rm 3} k}{8\pi^{\rm 2}B\mu_{\rm D}^{\rm 2}}\frac{e^{\left[hBJ(J+1)/kT_{\rm ex}\right]}}{J+1}\times\frac{T_{\rm ex}+hB/{\rm 3} k}{1-e^{(-h\nu/kT_{\rm ex})}}\int\tau_{v}dv},
\end{equation}

where $B$=55101.012 MHz is the rotational constant, $\mu_{\rm D}$=0.11 debyes is permanent dipole moment for $^{13}$CO, and $J$ is the rotational quantum number of the lower state in the observed transition \citep{Chackerian83}.
In this paper, we adopt a typical abundance ratio of [$^{12}$CO]/[$^{13}$CO] = 60 and [H$_{2}$]/[$^{12}$CO] = 10$^{4}$, considering the [$^{12}$CO]/[$^{13}$CO] ratio of $\sim$50 shown by \citet{Hawkins87} for the solar neighborhood and the ratio of $\sim$70 in the Galaxy shown by \citet{Penzias80}.
With these assumptions, the H$_2$ column densities are calculated and also presented in Table 6. The mean, maximum, minimum and median values
of column density of each dense core are also given in columns 2 to 5, respectively.
The $N_{\rm H_2}$ distributions, derived from PMO data, are presented in panel b of Figure 3.

The thermal velocity dispersion ($\sigma_{\rm th}$) and non-thermal
velocity dispersion ($\sigma_{\rm NT}$) are calculated as:
\begin{equation}
\it{\sigma_{\rm NT} = \left[\sigma_{\rm ^{\rm 13}CO}^{\rm 2}-\frac{kT_{\rm ex}}{m_{\rm ^{\rm 13}CO}}\right]^{\rm 1/2}},
\end{equation}
\begin{equation}
\it{\sigma_{\rm th} = \sqrt{\frac{kT_{\rm ex}}{m_{\rm H}\mu}}},
\end{equation}
where $m_{\rm ^{13}CO}$ is the mass of the $^{13}$CO molecule, $m_{\rm H}$ is the atomic hydrogen mass, and $\mu$ = 2.8 is the mean molecular weight of the gas \citep{Kauffmann08}. $\sigma_{^{13}CO}$ is the one-dimensional velocity dispersion, obtained from second moment maps. If the velocity dispersion is isotropic, the three-dimensional velocity dispersions are calculated using:
\begin{equation}
\it{\sigma_{\rm 3D} = \sqrt{{\rm 3}(\sigma_{\rm th}^{\rm 2}+\sigma_{\rm NT}^{\rm 2})}}
\end{equation}
The mean, maximum, minimum and median $\sigma_{\rm NT}$, $\sigma_{\rm th}$, and $\sigma_{\rm 3D}$
values of each dense core are also presented in Table 6.
The mean values of $\sigma_{\rm NT}$, derived from
$^{13}$CO, range from 0.36($\pm$0.09) km s$^{-1}$ to 0.73($\pm$0.07) km s$^{-1}$.
The mean values of $\sigma_{\rm th}$ range from 0.16($\pm$0.01) to 0.21($\pm$0.01) km s$^{-1}$.

If $^{13}$CO is optically thick, the $\sigma_{\rm NT}$ could be overestimated. Actually, $^{13}$CO is optically thick in PGCCs \citep{Wu12}.
The high spectral resolution (0.16 km~s$^{-1}$) in the PMO observations can also well resolve the C$^{18}$O line profiles.
Therefore, we also calculated $\sigma_{\rm NT}$ using C$^{18}$O data, and the $\sigma_{\rm NT}$(C$^{18}$O) values are presented in Table 8.
The $\sigma_{\rm NT}$(C$^{18}$O) values range from 0.19($\pm$0.01) km s$^{-1}$ to 0.71($\pm$0.21) km s$^{-1}$.
The Mach number ($\sigma_{\rm NT}$(C$^{18}$O)/$\sigma_{\rm th}$) can be calculated and listed in Table 8, where $\sigma_{\rm th}$ is the isothermal sound speed.
The Mach number of dense cores range from 1($\pm$0.08) to 3.8($\pm$1.2), with a mean value of 2.1($\pm$0.8).
Thus, we conclude that all the dense cores studied here have supersonic non-thermal motions.

For all of the dense cores, the Mach numbers are always greater than 1, and there is not much
difference in Mach number among starless cores, prestellar core candidates, and protostellar cores.
The large Mach numbers also indicate that the dense cores may be turbulence-dominated. The non-thermal velocity dispersions. however, could be also caused by bulk motions like infall or rotation.

The integrated intensity map of C$^{18}$O(1-0) are shown in panel c of Figure 3.
The C$^{18}$O map is very noisy (the average signal-to-noise ratio of C$^{18}$O mapping region is 4.1$\pm$1.2). Therefore, it is hard to derive accurate C$^{18}$O column density maps. Instead, we calculate core-averaged C$^{18}$O column densities from the core-averaged C$^{18}$O spectra.

If we assume that C$^{18}$O emission is optically thin and the excitation temperature is same as \emph{Herschel} dust temperature ($T_{\rm d}$), the $N_{\rm C^{18}O}$ (column density of C$^{18}$O) is calculated from \citep{Garden91}:
\begin{equation}
N_{C^{18}O}= 4.77\times 10^{13}\frac{(T_{\rm ex}+0.89)}{e^{(-5.27/T_{\rm ex})}}\int\frac{T_{\rm a}^{\ast}(\rm C^{18}O)}{\eta_{\rm b}}dv,
\end{equation}
where $T_{\rm a}^{\ast}$ is antenna temperature, and $\eta_{\rm b}$=0.6 is main beam efficiency of PMO telescope.

We applied the non-LTE code ``Radex" \citep{Van07} to fit C$^{18}$O parameters. The input parameters for Radex are volume density, kinetic temperature, and line width. We used volume density from \emph{Herschel} observations and assumed that the kinetic temperature equals to the dust temperature. The line widths were obtained from Gaussian fits to the PMO C$^{18}$O(1-0) spectra. The derived column density, excitation temperature and optical depth from Radex, and the LTE column density derived from PMO data are summarized in Table 7.

Figure 6 shows the correlation between the LTE C$^{18}$O column density and the non-LTE C$^{18}$O column density. The correlation is well fitted by a power-law function, and the fitted function is close to $N_{\rm C^{18}O}$(LTE) = $N_{\rm C^{18}O}$(non-LTE), which indicates that both the calculations are reasonable.

\begin{deluxetable}{cccccccccccccccccccccccccccccccccc}
\tabletypesize{\scriptsize} \tablecolumns{11}
\tablewidth{0pc} \setlength{\tabcolsep}{0.05in}
\tablecaption{The parameters derived from spectra of dense cores} \tablehead{
 Name &$V_{\rm lsr}$(12) &FWHM(12)      &$T_{\rm b}$(12) &$V_{\rm lsr}$(13) &FWHM(13)      &$T_{\rm b}$(13) &$V_{\rm lsr}$(18) &FWHM(18)      &$T_{\rm b}$(18)                \\
      &(km~s$^{-1}$) &(km~s$^{-1}$) &(K)         &(km~s$^{-1}$) &(km~s$^{-1}$) &(K)         &(km~s$^{-1}$) &(km~s$^{-1}$) &(K)                                                \\
}\startdata
G166.99-15.34-H1  &8.01($\pm$0.01)  &0.98($\pm$0.06)  &1.8($\pm$0.5)  &5.65($\pm$0.04)  &2.60($\pm$0.26)  &0.3($\pm$0.2)                                                      \\
                  &6.38($\pm$0.02)  &2.98($\pm$0.04)  &4.7($\pm$0.3)  &6.64($\pm$0.01)  &1.80($\pm$0.04)  &1.5($\pm$0.5)   &6.93($\pm$0.10)  &1.00($\pm$0.38)  &0.2($\pm$0.1)           \\
G167.23-15.32-H1  &4.42($\pm$0.04)  &1.23($\pm$0.12)  &3.1($\pm$0.6)  &4.23($\pm$0.24)  &2.05($\pm$0.53)  &0.4($\pm$0.3)                                                      \\
                  &6.92($\pm$0.04)  &2.97($\pm$0.10)  &5.8($\pm$0.4)  &6.58($\pm$0.03)  &1.54($\pm$0.08)  &2.8($\pm$0.3)   &6.65($\pm$0.04)  &0.53($\pm$0.13)  &1.0($\pm$0.3)           \\
G168.00-15.69-H1  &4.93($\pm$0.03)  &1.22($\pm$0.05)  &1.5($\pm$0.8)                                                                                                \\
                  &7.95($\pm$0.01)  &1.81($\pm$0.01)  &10.1($\pm$0.7) &7.83($\pm$0.01)  &0.92($\pm$0.01)  &6.4($\pm$0.2)   &7.79($\pm$0.01)  &0.55($\pm$0.02)  &1.8($\pm$0.1)           \\
G168.13-16.39-H1  &6.22($\pm$0.01)  &1.26($\pm$0.02)  &8.3($\pm$0.7)  &6.23($\pm$0.01)  &0.61($\pm$0.01)  &3.9($\pm$0.4)   &6.60($\pm$0.01)  &0.75($\pm$0.02)  &2.8($\pm$0.2)           \\
                  &7.46($\pm$0.01)  &0.92($\pm$0.02)  &7.2($\pm$0.7)  &7.00($\pm$0.01)  &0.90($\pm$0.01)  &5.5($\pm$0.3)                                                      \\
G168.13-16.39-H2  &3.60($\pm$0.06)  &1.57($\pm$0.13)  &2.0($\pm$0.6)                                                                                                \\
                  &6.81($\pm$0.02)  &2.09($\pm$0.04)  &8.6($\pm$0.5)  &6.84($\pm$0.01)  &1.24($\pm$0.01)  &6.6($\pm$0.3)   &6.84($\pm$0.02)  &0.88($\pm$0.04)  &2.1($\pm$0.2)           \\
G168.72-15.48-H1  &7.34($\pm$0.02)  &3.04($\pm$0.04)  &9.7($\pm$0.5)  &7.34($\pm$0.02)  &1.28($\pm$0.04)  &6.5($\pm$0.5)   &7.37($\pm$0.02)  &0.57($\pm$0.04)  &2.7($\pm$0.4)           \\
G168.72-15.48-H2  &7.40($\pm$0.01)  &2.74($\pm$0.03)  &10.3($\pm$0.4) &7.32($\pm$0.01)  &1.27($\pm$0.02)  &7.1($\pm$0.4)   &7.31($\pm$0.01)  &0.68($\pm$0.03)  &3.9($\pm$0.3)           \\
G169.43-16.17-H1  &5.24($\pm$0.02)  &1.80($\pm$0.04)  &7.4($\pm$0.8)  &5.35($\pm$0.02)  &1.42($\pm$0.03)  &5.0($\pm$0.8)   &5.53($\pm$0.03)  &1.30($\pm$0.07)  &1.4($\pm$0.2)           \\
                  &7.76($\pm$0.02)  &1.92($\pm$0.05)  &6.1($\pm$0.8)  &6.93($\pm$0.15)  &2.30($\pm$0.18)  &1.1($\pm$0.6)                                                      \\
G169.43-16.17-H2  &4.96($\pm$0.02)  &1.89($\pm$0.05)  &7.8($\pm$0.6)  &5.37($\pm$0.01)  &1.65($\pm$0.03)  &5.6($\pm$0.3)   &5.64($\pm$0.03)  &0.98($\pm$0.09)  &1.6($\pm$0.2)           \\
                  &7.56($\pm$0.02)  &1.63($\pm$0.05)  &7.9($\pm$0.7)  &7.52($\pm$0.03)  &1.29($\pm$0.08)  &2.0($\pm$0.3)                                                      \\
G169.76-16.15-H1  &5.21($\pm$0.16)  &1.67($\pm$0.16)  &7.1($\pm$0.9)  &5.52($\pm$0.01)  &1.76($\pm$0.03)  &5.2($\pm$0.4)   &5.69($\pm$0.02)  &1.28($\pm$0.06)  &1.9($\pm$0.2)           \\
                  &7.59($\pm$0.16)  &1.75($\pm$0.16)  &9.3($\pm$0.9)  &7.85($\pm$0.03)  &1.19($\pm$0.08)  &1.8($\pm$0.4)                                                      \\
G169.76-16.15-H2  &5.32($\pm$0.01)  &1.39($\pm$0.02)  &7.3($\pm$0.6)  &5.42($\pm$0.01)  &1.05($\pm$0.02)  &4.4($\pm$0.8)   &5.83($\pm$0.02)  &1.15($\pm$0.05)  &2.0($\pm$0.2)           \\
                  &7.56($\pm$0.01)  &1.94($\pm$0.03)  &9.2($\pm$0.6)  &6.71($\pm$0.03)  &2.18($\pm$0.05)  &3.4($\pm$0.6)                                                      \\
G169.76-16.15-H3  &5.40($\pm$0.01)  &1.39($\pm$0.01)  &7.6($\pm$0.6)  &5.40($\pm$0.01)  &0.88($\pm$0.02)  &4.5($\pm$0.5)   &5.52($\pm$0.01)  &0.59($\pm$0.02)  &2.2($\pm$0.2)           \\
                  &7.49($\pm$0.01)  &1.61($\pm$0.01)  &9.2($\pm$0.6)  &6.84($\pm$0.02)  &1.88($\pm$0.04)  &3.7($\pm$0.3)   &6.90($\pm$0.01)  &0.72($\pm$0.04)  &1.2($\pm$0.1)           \\
G169.76-16.15-H4  &5.48($\pm$0.01)  &1.38($\pm$0.02)  &7.6($\pm$0.5)  &5.45($\pm$0.01)  &0.67($\pm$0.02)  &2.8($\pm$0.5)   &5.59($\pm$0.03)  &0.58($\pm$0.08)  &0.6($\pm$0.2)           \\
                  &7.58($\pm$0.01)  &1.51($\pm$0.02)  &8.3($\pm$0.5)  &6.83($\pm$0.01)  &1.64($\pm$0.03)  &3.9($\pm$0.3)   &6.88($\pm$0.01)  &0.60($\pm$0.03)  &1.6($\pm$0.2)           \\
G169.76-16.15-H5  &5.57($\pm$0.01)  &1.28($\pm$0.03)  &7.8($\pm$0.6)  &5.59($\pm$0.02)  &0.77($\pm$0.03)  &3.8($\pm$0.5)   &5.71($\pm$0.06)  &0.89($\pm$0.13)  &0.6($\pm$0.2)           \\
                  &7.57($\pm$0.02)  &1.58($\pm$0.04)  &5.9($\pm$0.5)  &6.73($\pm$0.02)  &1.31($\pm$0.06)  &3.9($\pm$0.4)   &6.90($\pm$0.02)  &0.51($\pm$0.05)  &1.3($\pm$0.2)           \\
G170.00-16.14-H1  &5.47($\pm$0.02)  &1.28($\pm$0.05)  &7.6($\pm$0.7)  &5.70($\pm$0.12)  &1.25($\pm$0.18)  &2.8($\pm$0.9)                                                      \\
                  &7.36($\pm$0.03)  &1.56($\pm$0.07)  &5.4($\pm$0.6)  &6.78($\pm$0.08)  &1.06($\pm$0.11)  &3.6($\pm$0.9)   &6.59($\pm$0.08)  &1.11($\pm$0.13)  &0.7($\pm$0.3)           \\
G170.00-16.14-H2  &5.30($\pm$0.02)  &1.30($\pm$0.04)  &7.5($\pm$0.6)  &5.31($\pm$0.05)  &0.88($\pm$0.11)  &1.8($\pm$0.4)                                                      \\
                  &7.40($\pm$0.04)  &2.12($\pm$0.09)  &5.2($\pm$0.5)  &6.64($\pm$0.03)  &1.56($\pm$0.08)  &4.4($\pm$0.4)   &6.61($\pm$0.05)  &1.11($\pm$0.10)  &0.9($\pm$0.2)           \\
G170.13-16.06-H1  &5.94($\pm$0.02)  &1.48($\pm$0.04)  &7.2($\pm$0.5)  &6.00($\pm$0.02)  &0.96($\pm$0.03)  &4.5($\pm$0.4)   &5.97($\pm$0.06)  &0.62($\pm$0.17)  &0.6($\pm$0.3)           \\
                  &7.38($\pm$0.02)  &1.00($\pm$0.04)  &4.2($\pm$0.6)  &6.96($\pm$0.02)  &0.91($\pm$0.03)  &4.1($\pm$0.4)   &6.85($\pm$0.03)  &0.80($\pm$0.07)  &1.4($\pm$0.3)           \\
G170.26-16.02-H1  &6.23($\pm$0.02)  &2.36($\pm$0.04)  &6.7($\pm$0.7)  &6.36($\pm$0.01)  &1.66($\pm$0.02)  &5.1($\pm$0.4)   &6.64($\pm$0.04)  &1.41($\pm$0.09)  &1.1($\pm$0.2)           \\
G170.26-16.02-H2  &5.99($\pm$0.05)  &1.22($\pm$0.08)  &6.7($\pm$1.2)  &6.51($\pm$0.01)  &1.24($\pm$0.01)  &5.5($\pm$0.2)   &6.56($\pm$0.01)  &0.83($\pm$0.04)  &1.6($\pm$0.1)           \\
                  &7.17($\pm$0.06)  &1.13($\pm$0.09)  &4.9($\pm$1.2)                                                                                                \\
G170.83-15.90-H1  &6.61($\pm$0.02)  &1.46($\pm$0.04)  &6.2($\pm$0.4)  &6.53($\pm$0.01)  &1.06($\pm$0.02)  &5.3($\pm$0.3)   &6.57($\pm$0.04)  &1.17($\pm$0.12)  &1.3($\pm$0.3)           \\
G170.83-15.90-H2  &6.67($\pm$0.02)  &1.73($\pm$0.04)  &6.2($\pm$0.5)  &6.66($\pm$0.01)  &1.18($\pm$0.02)  &5.0($\pm$0.2)   &6.79($\pm$0.07)  &0.95($\pm$0.17)  &1.1($\pm$0.3)           \\
G170.99-15.81-H1  &5.56($\pm$0.10)  &1.06($\pm$0.16)  &2.8($\pm$0.4)                                                                                                \\
                  &6.78($\pm$0.06)  &1.13($\pm$0.18)  &4.3($\pm$0.4)  &6.50($\pm$0.01)  &1.16($\pm$0.03)  &4.8($\pm$0.2)   &6.51($\pm$0.02)  &0.64($\pm$0.05)  &2.2($\pm$0.3)           \\
                  &8.07($\pm$0.07)  &0.71($\pm$0.17)  &1.7($\pm$0.8)                                                                                                \\
G171.49-14.90-H1  &5.95($\pm$0.16)  &1.29($\pm$0.16)  &7.0($\pm$0.6)  &6.52($\pm$0.02)  &0.92($\pm$0.04)  &5.3($\pm$0.5)   &6.55($\pm$0.02)  &0.47($\pm$0.04)  &2.4($\pm$0.4)           \\
                  &7.49($\pm$0.16)  &1.31($\pm$0.16)  &8.4($\pm$0.7)  &7.58($\pm$0.03)  &0.85($\pm$0.05)  &3.2($\pm$0.5)                                                      \\
G171.80-15.32-H1  &6.90($\pm$0.26)  &1.77($\pm$0.04)  &7.4($\pm$0.5)  &6.99($\pm$0.01)  &0.96($\pm$0.02)  &5.7($\pm$0.3)   &6.96($\pm$0.06)  &1.22($\pm$0.13)  &0.9($\pm$0.2)           \\
G171.80-15.32-H2  &9.06($\pm$0.02)  &2.17($\pm$0.05)  &7.4($\pm$0.5)  &7.23($\pm$0.02)  &1.29($\pm$0.04)  &5.2($\pm$0.3)   &7.38($\pm$0.13)  &1.67($\pm$0.30)  &0.7($\pm$0.3)           \\
G171.80-15.32-H3  &7.04($\pm$0.02)  &1.96($\pm$0.04)  &6.7($\pm$0.4)  &7.17($\pm$0.01)  &1.01($\pm$0.02)  &5.1($\pm$0.3)   &7.12($\pm$0.11)  &1.44($\pm$0.27)  &0.5($\pm$0.3)           \\
G171.91-15.65-H1  &6.59($\pm$0.02)  &2.36($\pm$0.04)  &7.4($\pm$0.4)  &6.61($\pm$0.01)  &1.25($\pm$0.03)  &6.2($\pm$0.3)   &6.56($\pm$0.04)  &0.90($\pm$0.09)  &1.4($\pm$0.3)           \\
G171.91-15.65-H2  &6.54($\pm$0.02)  &2.12($\pm$0.05)  &7.3($\pm$0.4)  &6.61($\pm$0.01)  &1.14($\pm$0.02)  &6.5($\pm$0.3)   &6.64($\pm$0.03)  &0.78($\pm$0.06)  &1.9($\pm$0.3)           \\
G171.91-15.65-H3  &6.52($\pm$0.02)  &1.97($\pm$0.05)  &7.8($\pm$0.4)  &6.60($\pm$0.01)  &1.06($\pm$0.02)  &6.1($\pm$0.3)   &6.66($\pm$0.02)  &0.69($\pm$0.06)  &1.7($\pm$0.3)           \\
G172.06-15.21-H1  &6.59($\pm$0.01)  &1.72($\pm$0.02)  &9.3($\pm$0.2)  &6.77($\pm$0.01)  &1.01($\pm$0.02)  &5.1($\pm$0.2)   &6.81($\pm$0.01)  &0.57($\pm$0.03)  &1.8($\pm$0.2)           \\
\enddata
\tablecomments{All parameters are derived from spectra of dense cores, which averaged over identified core's region. For some dense cores, there are multiple velocity components, But there are some velocity components that don't have $^{13}$CO and C$^{18}$O emission. The spectral plot can be seen in Figure 5.}
\end{deluxetable}

\begin{deluxetable}{cccccccccccccccccccccccc} 
\tabletypesize{\tiny} \tablecolumns{24}
\tablewidth{0pc} \rotate \setlength{\tabcolsep}{0.04in}
\tablecaption{The parameters of dense core derived from CO gas emission} \tablehead{
 Name  & \multicolumn{4}{c}{$N_{\rm H_{2}}$(PMO)}  &\multicolumn{4}{c}{$T_{\rm ex}$} &$\tau$ ($^{13}$CO) &\multicolumn{4}{c}{$\sigma_{\rm th}$} &\multicolumn{4}{c}{$\sigma_{\rm NT}$ ($^{13}$CO)} &\multicolumn{4}{c}{$\sigma_{\rm 3D}$ ($^{13}$CO)}\\
\cline{2-22}
 & mean & max & min & median & mean & max & min & median & & mean & max & min & median & mean & max & min & median & mean & max & min & median\\
\cline{2-22} \\
 &\multicolumn{4}{c}{(10$^{21}$ cm$^{-2}$)} &\multicolumn{4}{c}{(K)} & &\multicolumn{4}{c}{(km~s$^{-1}$)}   &\multicolumn{4}{c}{(km~s$^{-1}$)} &\multicolumn{4}{c}{(km~s$^{-1}$)}
}\startdata
G166.99-15.34-H1  &2.1($\pm$0.2) &2.4 &1.7 &2.1 &8.5($\pm$0.3)  &9.0  &7.8  &8.4  &0.34($\pm$0.03) &0.15($\pm$0.003) &0.16 &0.14 &0.15 &0.56($\pm$0.11) &0.83  &0.22  &0.62 &1.02($\pm$0.29) &1.45  &0.45 &1.11   \\
G167.23-15.32-H1  &3.0($\pm$0.4) &3.7 &2.2 &2.9 &10.7($\pm$0.8) &12.5 &9.2  &10.8 &0.48($\pm$0.07) &0.16($\pm$0.007) &0.17 &0.15 &0.16 &0.65($\pm$0.21) &1.26  &0.60  &0.96 &1.17($\pm$0.34) &2.20  &1.07 &1.69   \\
G168.00-15.69-H1  &4.8($\pm$0.3) &5.5 &3.9 &4.8 &15.1($\pm$0.7) &16.8 &13.9 &14.9 &0.78($\pm$0.07) &0.21($\pm$0.005) &0.23 &0.21 &0.21 &0.48($\pm$0.06) &0.62  &0.38  &0.49 &0.91($\pm$0.09) &1.14  &0.75 &0.92   \\
G168.13-16.39-H1  &3.2($\pm$0.3) &4.1 &3.0 &3.5 &12.1($\pm$0.4) &13.4 &11.3 &12.0 &0.58($\pm$0.04) &0.19($\pm$0.004) &0.20 &0.19 &0.19 &0.37($\pm$0.04) &0.44  &0.27  &0.38 &0.73($\pm$0.06) &0.83  &0.58 &0.73   \\
G168.13-16.39-H2  &3.5($\pm$0.3) &4.1 &3.1 &3.5 &12.5($\pm$0.7) &14.1 &11.4 &12.3 &1.30($\pm$0.20) &0.19($\pm$0.005) &0.21 &0.19 &0.19 &0.36($\pm$0.06) &0.48  &0.26  &0.37 &0.70($\pm$0.09) &0.91  &0.55 &0.72   \\
G168.72-15.48-H1  &6.8($\pm$0.5) &8.1 &5.6 &6.9 &14.4($\pm$0.4) &16.6 &12.3 &14.3 &1.03($\pm$0.07) &0.21($\pm$0.003) &0.73 &0.19 &0.21 &0.73($\pm$0.07) &0.89  &0.37  &0.75 &1.32($\pm$0.12) &1.59  &0.73 &1.35   \\
G168.72-15.48-H2  &6.8($\pm$0.6) &7.8 &5.9 &6.9 &14.9($\pm$0.7) &16.6 &13.7 &14.9 &0.84($\pm$0.07) &0.21($\pm$0.005) &0.22 &0.20 &0.21 &0.67($\pm$0.09) &0.82  &0.43  &0.67 &1.21($\pm$0.15) &1.47  &0.84 &1.21   \\
G169.43-16.17-H1  &6.1($\pm$0.4) &7.0 &5.3 &6.0 &11.0($\pm$0.4) &12.1 &10.1 &11.0 &1.07($\pm$0.10) &0.18($\pm$0.003) &0.19 &0.18 &0.18 &0.58($\pm$0.05) &0.66  &0.46  &0.59 &1.05($\pm$0.08) &1.19  &0.85 &1.07   \\
G169.43-16.17-H2  &1.8($\pm$0.3) &2.1 &1.1 &1.8 &11.9($\pm$0.8) &13.7 &10.6 &11.8 &1.07($\pm$0.17) &0.19($\pm$0.006) &0.19 &0.18 &0.19 &0.40($\pm$0.16) &0.70  &0.46  &0.65 &0.77($\pm$0.24) &1.27  &1.00 &1.18   \\
G169.76-16.15-H1  &5.3($\pm$0.4) &5.9 &4.5 &5.2 &11.2($\pm$0.6) &12.7 &10.1 &11.0 &1.09($\pm$0.15) &0.18($\pm$0.005) &0.20 &0.18 &0.18 &0.53($\pm$0.04) &0.60  &0.47  &0.53 &0.98($\pm$0.07) &1.09  &0.87 &0.97   \\
G169.76-16.15-H2  &4.8($\pm$0.2) &5.2 &4.4 &4.9 &11.0($\pm$0.6) &12.4 &10.1 &10.8 &0.85($\pm$0.10) &0.18($\pm$0.005) &0.19 &0.18 &0.18 &0.47($\pm$0.04) &0.56  &0.41  &0.47 &0.88($\pm$0.06) &1.01  &0.78 &0.86   \\
G169.76-16.15-H3  &3.5($\pm$0.3) &4.1 &2.9 &3.5 &12.9($\pm$0.9) &14.4 &10.6 &13.0 &0.48($\pm$0.06) &0.20($\pm$0.007) &0.19 &0.18 &0.18 &0.51($\pm$0.07) &0.54  &0.38  &0.47 &0.95($\pm$0.11) &0.99  &0.74 &0.87   \\
G169.76-16.15-H4  &3.2($\pm$0.2) &3.8 &2.8 &3.2 &11.6($\pm$0.7) &12.5 &10.2 &11.8 &0.66($\pm$0.08) &0.19($\pm$0.006) &0.20 &0.18 &0.19 &0.49($\pm$0.11) &0.59  &0.27  &0.48 &0.91($\pm$0.18) &1.06  &0.57 &0.89   \\
G169.76-16.15-H5  &2.5($\pm$0.3) &3.0 &2.0 &2.6 &10.2($\pm$0.7) &11.2 &9.4  &10.3 &0.81($\pm$0.13) &0.18($\pm$0.005) &0.20 &0.18 &0.19 &0.45($\pm$0.14) &0.57  &0.37  &0.45 &0.84($\pm$0.23) &1.05  &0.72 &0.85   \\
G170.00-16.14-H1  &3.7($\pm$0.3) &4.0 &3.2 &3.8 &11.4($\pm$0.6) &12.4 &10.4 &11.7 &0.59($\pm$0.06) &0.19($\pm$0.005) &0.20 &0.18 &0.19 &0.52($\pm$0.18) &0.69  &0.16  &0.64 &0.97($\pm$0.28) &1.24  &0.42 &1.16   \\
G170.00-16.14-H2  &3.8($\pm$0.7) &4.7 &2.4 &3.7 &11.6($\pm$0.8) &13.8 &10.7 &11.4 &0.76($\pm$0.11) &0.19($\pm$0.007) &0.21 &0.18 &0.19 &0.59($\pm$0.10) &0.67  &0.31  &0.62 &1.07($\pm$0.16) &1.20  &0.62 &1.12   \\
G170.13-16.06-H1  &5.0($\pm$0.4) &5.9 &3.9 &5.0 &11.4($\pm$0.7) &12.8 &10.1 &11.5 &0.72($\pm$0.09) &0.19($\pm$0.006) &0.20 &0.18 &0.19 &0.55($\pm$0.04) &0.63  &0.44  &0.56 &1.01($\pm$0.07) &1.13  &0.82 &1.02   \\
G170.26-16.02-H1  &3.9($\pm$0.3) &4.2 &3.6 &3.8 &11.4($\pm$0.6) &12.3 &10.6 &11.4 &1.01($\pm$0.12) &0.18($\pm$0.005) &0.22 &0.16 &0.18 &0.43($\pm$0.02) &0.77  &0.30  &0.43 &0.82($\pm$0.08) &1.38  &0.60 &0.82   \\
G170.26-16.02-H2  &4.2($\pm$0.1) &4.5 &3.9 &4.2 &10.6($\pm$0.5) &11.4 &9.7  &10.7 &1.42($\pm$0.21) &0.18($\pm$0.004) &0.19 &0.17 &0.18 &0.46($\pm$0.05) &0.54  &0.32  &0.48 &0.86($\pm$0.09) &0.98  &0.63 &0.88   \\
G170.83-15.90-H1  &1.7($\pm$0.4) &2.4 &9.3 &1.7 &9.6($\pm$0.8)  &11.4 &8.5  &9.5  &1.81($\pm$0.65) &0.17($\pm$0.007) &0.19 &0.16 &0.17 &0.39($\pm$0.14) &0.58  &0.07  &0.41 &0.74($\pm$0.22) &1.05  &0.31 &0.76   \\
G170.83-15.90-H2  &1.3($\pm$0.4) &2.3 &3.2 &1.3 &10.2($\pm$0.9) &12.4 &8.1  &10.2 &1.28($\pm$0.32) &0.18($\pm$0.008) &0.19 &0.16 &0.18 &0.37($\pm$0.11) &0.69  &0.11  &0.39 &0.72($\pm$0.18) &1.24  &0.34 &0.73   \\
G170.99-15.81-H1  &3.8($\pm$0.9) &5.8 &2.0 &3.7 &8.9($\pm$1.0)  &11.6 &7.1  &8.8  &1.89($\pm$1.01) &0.16($\pm$0.010) &0.19 &0.15 &0.16 &0.58($\pm$0.18) &0.94  &0.14  &0.61 &1.05($\pm$0.30) &1.65  &0.38 &1.09   \\
G171.49-14.90-H1  &4.5($\pm$0.5) &5.1 &3.8 &4.5 &12.4($\pm$0.8) &14.0 &11.6 &12.2 &0.88($\pm$0.12) &0.19($\pm$0.006) &0.20 &0.18 &0.18 &0.51($\pm$0.08) &0.54  &0.14  &0.38 &0.95($\pm$0.13) &0.99  &0.39 &0.73   \\
G171.80-15.32-H1  &3.9($\pm$0.5) &4.5 &3.3 &3.8 &12.1($\pm$0.7) &13.6 &11.0 &12.0 &1.05($\pm$0.14) &0.19($\pm$0.005) &0.20 &0.18 &0.19 &0.43($\pm$0.13) &0.66  &0.20  &0.44 &0.81($\pm$0.20) &1.18  &0.48 &0.83   \\
G171.80-15.32-H2  &4.5($\pm$0.3) &5.0 &4.2 &4.5 &11.4($\pm$0.6) &12.9 &10.6 &11.5 &1.02($\pm$0.13) &0.19($\pm$0.005) &0.20 &0.18 &0.19 &0.65($\pm$0.06) &0.75  &0.46  &0.66 &1.16($\pm$0.11) &1.32  &0.86 &1.19   \\
G171.80-15.32-H3  &3.7($\pm$0.3) &4.2 &3.1 &3.6 &10.8($\pm$0.4) &11.6 &10.3 &10.7 &1.14($\pm$0.10) &0.18($\pm$0.003) &0.19 &0.18 &0.18 &0.66($\pm$0.07) &0.76  &0.53  &0.66 &1.19($\pm$0.11) &1.36  &0.97 &1.18   \\
G171.91-15.65-H1  &4.8($\pm$0.3) &5.3 &4.3 &4.6 &11.4($\pm$0.5) &12.3 &10.8 &11.4 &1.49($\pm$0.19) &0.19($\pm$0.004) &0.19 &0.18 &0.19 &0.42($\pm$0.09) &0.52  &0.27  &0.43 &0.80($\pm$0.13) &0.96  &0.57 &0.81   \\
G171.91-15.65-H2  &4.6($\pm$0.5) &5.3 &3.6 &4.7 &11.3($\pm$0.4) &12.1 &10.9 &11.3 &1.64($\pm$0.19) &0.19($\pm$0.003) &0.19 &0.18 &0.19 &0.43($\pm$0.08) &0.53  &0.27  &0.43 &0.80($\pm$0.14) &0.97  &0.13 &0.80   \\
G171.91-15.65-H3  &4.4($\pm$0.2) &4.5 &3.8 &4.2 &12.2($\pm$0.5) &13.0 &11.0 &12.1 &1.16($\pm$0.12) &0.19($\pm$0.004) &0.20 &0.18 &0.19 &0.36($\pm$0.09) &0.48  &0.23  &0.40 &0.71($\pm$0.14) &0.89  &0.52 &0.77   \\
G172.06-15.21-H1  &3.7($\pm$0.3) &4.3 &3.2 &3.7 &13.3($\pm$0.4) &13.9 &12.4 &13.2 &0.71($\pm$0.05) &0.20($\pm$0.003) &0.21 &0.20 &0.20 &0.41($\pm$0.06) &0.52  &0.31  &0.41 &0.79($\pm$0.09) &0.97  &0.64 &0.79   \\
\enddata
\tablecomments{All of this parameters are derived from the PMO observational data. Excitation temperature is calculated by $^{12}$CO data, $N_{H_{2}}$(PMO), $\sigma_{\rm NT}$, and $\sigma_{\rm 3D}$ are derived from $^{13}$CO images.}
\end{deluxetable}

\begin{deluxetable}{cccccccccccccccccccccccccccccccccc}[h!] 
\tabletypesize{\scriptsize} \tablecolumns{9}
\tablewidth{0pc} \setlength{\tabcolsep}{0.05in}
\tablecaption{The parameters of C$^{18}$O molecules }
\tablehead{
 Name &$N_{\rm C^{18}O}$(LTE)       &C$^{18}$O abundance(LTE)      &$f_{\rm D}$   &$\tau$ (C$^{18}$O)  &$T_{\rm ex}$(non-LTE)(C$^{18}$O) &$N_{\rm C^{18}O}$(non-LTE)            \\
      &(10$^{14}$ cm$^{-2}$)        &(10$^{-7}$)                   &              &                    &(K)                              &(10$^{15}$ cm$^{-2}$)                 \\
}
\startdata
G166.99-15.34-H1     &1.59($\pm$0.22)   &0.44($\pm$0.18)   &6.59($\pm$1.75) &0.02 &13.99   &0.15          \\
G167.23-15.32-H1     &4.92($\pm$0.40)   &0.66($\pm$0.26)   &4.36($\pm$1.16) &0.10 &13.02   &0.18          \\
G168.00-15.69-H1     &9.92($\pm$0.11)   &1.69($\pm$0.42)   &1.69($\pm$0.45) &0.2. &12.31   &1.02          \\
G168.13-16.39-H1     &19.63($\pm$0.17)  &1.28($\pm$0.34)   &2.24($\pm$0.59) &0.38 &12.06   &2.54          \\
G168.13-16.39-H2     &17.16($\pm$0.15)  &1.29($\pm$0.15)   &2.22($\pm$0.59) &0.28 &11.80   &2.02          \\
G168.72-15.48-H1     &17.62($\pm$0.19)  &1.26($\pm$0.21)   &2.28($\pm$0.61) &0.37 &12.08   &2.24          \\
G168.72-15.48-H2     &22.32($\pm$1.55)  &0.91($\pm$0.29)   &3.17($\pm$0.84) &0.50 &13.31   &3.11          \\
G169.43-16.17-H1     &16.88($\pm$0.24)  &1.61($\pm$0.38)   &1.78($\pm$0.47) &0.17 &12.25   &1.83          \\
G169.43-16.17-H2     &14.58($\pm$0.36)  &1.48($\pm$0.31)   &1.95($\pm$0.52) &0.20 &12.12   &1.66          \\
G169.76-16.15-H1     &22.89($\pm$0.23)  &2.35($\pm$0.31)   &1.22($\pm$0.32) &0.24 &12.19   &2.69          \\
G169.76-16.15-H2     &22.05($\pm$0.10)  &2.31($\pm$0.23)   &1.24($\pm$0.33) &0.24 &12.53   &2.54          \\
G169.76-16.15-H3     &8.42($\pm$0.15)   &0.80($\pm$0.21)   &3.58($\pm$0.95) &0.14 &12.43   &0.96          \\
G169.76-16.15-H4     &9.23($\pm$0.15)   &0.83($\pm$0.19)   &3.46($\pm$0.92) &0.21 &12.05   &1.10          \\
G169.76-16.15-H5     &6.25($\pm$0.16)   &0.55($\pm$0.18)   &5.20($\pm$1.38) &0.16 &12.20   &0.74          \\
G170.00-16.14-H1     &7.56($\pm$0.29)   &0.61($\pm$0.17)   &4.70($\pm$1.25) &0.10 &11.25   &0.89          \\
G170.00-16.14-H2     &9.24($\pm$0.66)   &0.74($\pm$0.30)   &3.89($\pm$1.03) &0.10 &12.38   &1.08          \\
G170.13-16.06-H1     &10.43($\pm$0.31)  &0.90($\pm$0.23)   &3.18($\pm$0.84) &0.18 &11.79   &1.18          \\
G170.26-16.02-H1     &14.64($\pm$0.34)  &1.21($\pm$0.28)   &1.95($\pm$0.52) &0.14 &11.57   &1.74          \\
G170.26-16.02-H2     &1.22($\pm$0.22)   &0.99($\pm$0.32)   &2.91($\pm$0.77) &0.22 &11.23   &1.52          \\
G170.83-15.90-H1     &15.08($\pm$0.41)  &2.87($\pm$0.76)   &1.00($\pm$0.27) &0.16 &12.55   &1.59          \\
G170.83-15.90-H2     &10.04($\pm$0.49)  &1.86($\pm$0.36)   &1.54($\pm$0.41) &0.13 &12.66   &1.09          \\
G170.99-15.81-H1     &13.12($\pm$0.28)  &1.61($\pm$0.40)   &1.79($\pm$0.47) &0.30 &11.72   &1.48          \\
G171.49-14.90-H1     &10.21($\pm$0.34)  &0.48($\pm$0.17)   &5.97($\pm$1.58) &0.36 &11.35   &1.44          \\
G171.80-15.32-H1     &9.82($\pm$0.41)   &0.66($\pm$0.25)   &4.37($\pm$1.16) &0.12 &11.24   &1.17          \\
G171.80-15.32-H2     &10.82($\pm$1.44)  &1.21($\pm$0.37)   &2.36($\pm$0.63) &0.08 &12.76   &1.23          \\
G171.80-15.32-H3     &11.82($\pm$0.29)  &0.88($\pm$0.11)   &3.25($\pm$0.86) &0.06 &11.37   &0.72          \\
G171.91-15.65-H1     &12.82($\pm$0.51)  &1.64($\pm$0.60)   &1.75($\pm$0.46) &0.14 &14.10   &1.48          \\
G171.91-15.65-H2     &13.82($\pm$0.66)  &1.42($\pm$0.68)   &2.03($\pm$0.54) &0.21 &13.47   &1.80          \\
G171.91-15.65-H3     &14.82($\pm$0.13)  &1.80($\pm$0.17)   &1.60($\pm$0.42) &0.20 &12.69   &1.26          \\
G172.06-15.21-H1     &15.82($\pm$0.17)  &1.83($\pm$0.35)   &1.57($\pm$0.42) &0.22 &12.39   &1.13          \\
\enddata
\tablecomments{The LTE parameters are derived from C$^{18}$O spectra. The non-LTE parameters and $\tau$(C$^{18}$O) are derived from the ``Radex'' software. The depletion factor ($f_{D}$) is calculated by $N_{\rm C^{18}O}$(LTE)}
\end{deluxetable}

\begin{deluxetable}{cccccccccccccccccccccccccccccccccccc} 
\tabletypesize{\scriptsize} \tablecolumns{12}
\tablewidth{0pc} \setlength{\tabcolsep}{0.05in}
\tablecaption{The parameters of the dense cores} \tablehead{
 Name & $V_{\rm lsr}$(C$^{18}$O)  &$M_{\rm vir}$ & $\alpha$ &Jeans length &$\sigma_{\rm NT}$ (C$^{18}$O) &Mach number\tablenotemark{a}  \\
 &(km~s$^{-1}$)  &\colhead{(M$_{\sun}$)} & &(pc) &\colhead{(km~s$^{-1}$)}
}\startdata
G166.99-15.34-H1    &6.93($\pm$0.10)       &23.49($\pm$17.76) &6.50($\pm$5.20)  &0.375($\pm$0.011) &0.43($\pm$0.32)   &2.66($\pm$2.05)     \\
G167.23-15.32-H1    &6.65($\pm$0.04)       &6.48($\pm$3.06)   &0.97($\pm$0.49)   &0.249($\pm$0.007) &0.22($\pm$0.03)  &1.22($\pm$0.20)     \\
G168.00-15.69-H1    &7.79($\pm$0.01)       &16.82($\pm$1.21)  &0.58($\pm$0.04)   &0.440($\pm$0.010) &0.23($\pm$0.01)  &1.04($\pm$0.05)     \\
G168.13-16.39-H1    &6.60($\pm$0.01)       &20.20($\pm$1.02)  &0.58($\pm$0.03)   &0.207($\pm$0.006) &0.32($\pm$0.01)  &1.65($\pm$0.07)     \\
G168.13-16.39-H2    &6.84($\pm$0.02)       &35.58($\pm$3.31)  &0.98($\pm$0.09)   &0.244($\pm$0.004) &0.37($\pm$0.02)  &1.92($\pm$0.14)     \\
G168.72-15.48-H1    &7.37($\pm$0.02)       &16.66($\pm$1.22)  &0.63($\pm$0.05)   &0.214($\pm$0.005) &0.28($\pm$0.01)  &1.37($\pm$0.05)     \\
G168.72-15.48-H2    &7.31($\pm$0.01)       &9.49($\pm$1.27)   &0.25($\pm$0.03)   &0.164($\pm$0.013) &0.24($\pm$0.01)  &1.11($\pm$0.06)     \\
G169.43-16.17-H1    &5.53($\pm$0.03)       &74.28($\pm$7.64)  &2.45($\pm$0.26)   &0.275($\pm$0.006) &0.56($\pm$0.15)  &3.04($\pm$0.89)     \\
G169.43-16.17-H2    &5.64($\pm$0.03)       &28.21($\pm$5.30)  &2.28($\pm$0.45)   &0.231($\pm$0.005) &0.42($\pm$0.05)  &2.19($\pm$0.31)     \\
G169.76-16.15-H1    &5.69($\pm$0.02)       &37.45($\pm$3.46)  &5.78($\pm$0.62)   &0.205($\pm$0.004) &0.55($\pm$0.19)  &2.96($\pm$1.10)     \\
G169.76-16.15-H2    &5.83($\pm$0.02)       &32.81($\pm$2.56)  &4.64($\pm$0.42)   &0.222($\pm$0.002) &0.49($\pm$0.04)  &2.69($\pm$0.30)     \\
G169.76-16.15-H3    &6.90($\pm$0.01)       &11.85($\pm$1.15)  &1.26($\pm$0.14)   &0.202($\pm$0.005) &0.30($\pm$0.01)  &1.53($\pm$0.11)     \\
G169.76-16.15-H4    &6.88($\pm$0.01)       &7.97($\pm$0.78)   &0.91($\pm$0.10)   &0.189($\pm$0.005) &0.25($\pm$0.01)  &1.33($\pm$0.09)     \\
G169.76-16.15-H5    &6.90($\pm$0.02)       &3.86($\pm$0.73)   &0.87($\pm$0.20)   &0.157($\pm$0.004) &0.21($\pm$0.01)  &1.19($\pm$0.10)     \\
G170.00-16.14-H1    &6.59($\pm$0.08)       &16.00($\pm$3.62)  &4.97($\pm$1.37)   &0.128($\pm$0.004) &0.47($\pm$0.09)  &2.54($\pm$0.56)     \\
G170.00-16.14-H2    &6.61($\pm$0.05)       &16.68($\pm$3.01)  &3.66($\pm$0.81)   &0.143($\pm$0.009) &0.47($\pm$0.07)  &2.51($\pm$0.46)     \\
G170.13-16.06-H1    &6.85($\pm$0.03)       &25.16($\pm$4.34)  &0.94($\pm$0.17)   &0.240($\pm$0.007) &0.34($\pm$0.03)  &1.82($\pm$0.19)     \\
G170.26-16.02-H1    &6.64($\pm$0.04)       &20.28($\pm$2.53)  &15.08($\pm$3.43)  &0.146($\pm$0.004) &0.60($\pm$0.10)  &3.28($\pm$0.63)     \\
G170.26-16.02-H2    &6.56($\pm$0.01)       &12.67($\pm$1.07)  &1.63($\pm$0.17)   &0.153($\pm$0.006) &0.35($\pm$0.02)  &1.97($\pm$0.13)     \\
G170.83-15.90-H1    &6.57($\pm$0.04)       &35.56($\pm$7.05)  &7.09($\pm$1.66)   &0.305($\pm$0.007) &0.50($\pm$0.23)  &2.92($\pm$1.48)     \\
G170.83-15.90-H2    &6.79($\pm$0.07)       &18.27($\pm$6.41)  &4.53($\pm$1.68)   &0.267($\pm$0.006) &0.40($\pm$0.09)  &2.30($\pm$0.61)     \\
G170.99-15.81-H1    &6.51($\pm$0.02)       &26.87($\pm$3.84)  &0.45($\pm$0.07)   &0.377($\pm$0.009) &0.27($\pm$0.02)  &1.66($\pm$0.20)     \\
G171.49-14.90-H1    &6.55($\pm$0.02)       &3.60($\pm$0.65)   &0.36($\pm$0.07)   &0.111($\pm$0.004) &0.19($\pm$0.01)  &1.00($\pm$0.08)     \\
G171.80-15.32-H1    &6.96($\pm$0.06)       &25.72($\pm$5.42)  &3.27($\pm$0.77)   &0.134($\pm$0.005) &0.52($\pm$0.12)  &2.72($\pm$0.72)     \\
G171.80-15.32-H2    &7.38($\pm$0.13)       &35.66($\pm$12.67) &17.46($\pm$7.27)  &0.167($\pm$0.010) &0.71($\pm$0.21)  &3.83($\pm$1.20)     \\
G171.80-15.32-H3    &7.12($\pm$0.11)       &50.51($\pm$19.20) &4.47($\pm$1.75)   &0.169($\pm$0.004) &0.62($\pm$0.25)  &3.40($\pm$1.42)     \\
G171.91-15.65-H1    &6.56($\pm$0.04)       &7.50($\pm$1.45)   &6.17($\pm$2.15)   &0.169($\pm$0.005) &0.38($\pm$0.04)  &2.06($\pm$0.24)     \\
G171.91-15.65-H2    &6.64($\pm$0.03)       &6.40($\pm$0.96)   &2.56($\pm$0.58)   &0.157($\pm$0.007) &0.33($\pm$0.02)  &1.76($\pm$0.13)     \\
G171.91-15.65-H3    &6.66($\pm$0.02)       &9.61($\pm$1.53)   &2.27($\pm$0.41)   &0.218($\pm$0.002) &0.29($\pm$0.02)  &1.50($\pm$0.11)     \\
G172.06-15.21-H1    &6.81($\pm$0.01)       &7.30($\pm$0.79)   &0.94($\pm$0.12)   &0.219($\pm$0.006) &0.24($\pm$0.01)  &1.18($\pm$0.05)     \\
\enddata
\tablenotetext{a}{The Mach numbers are derived from $\sigma_{\rm NT}$(C$^{18}$O)/$\sigma_{\rm th}$}.
\tablecomments{$\alpha$ is alculated by $M_{\rm vir}$/$M_{\rm core}$, and the other parameters are derived from C$^{18}$O data.}
\end{deluxetable}

\begin{figure} 
\gridline{\fig{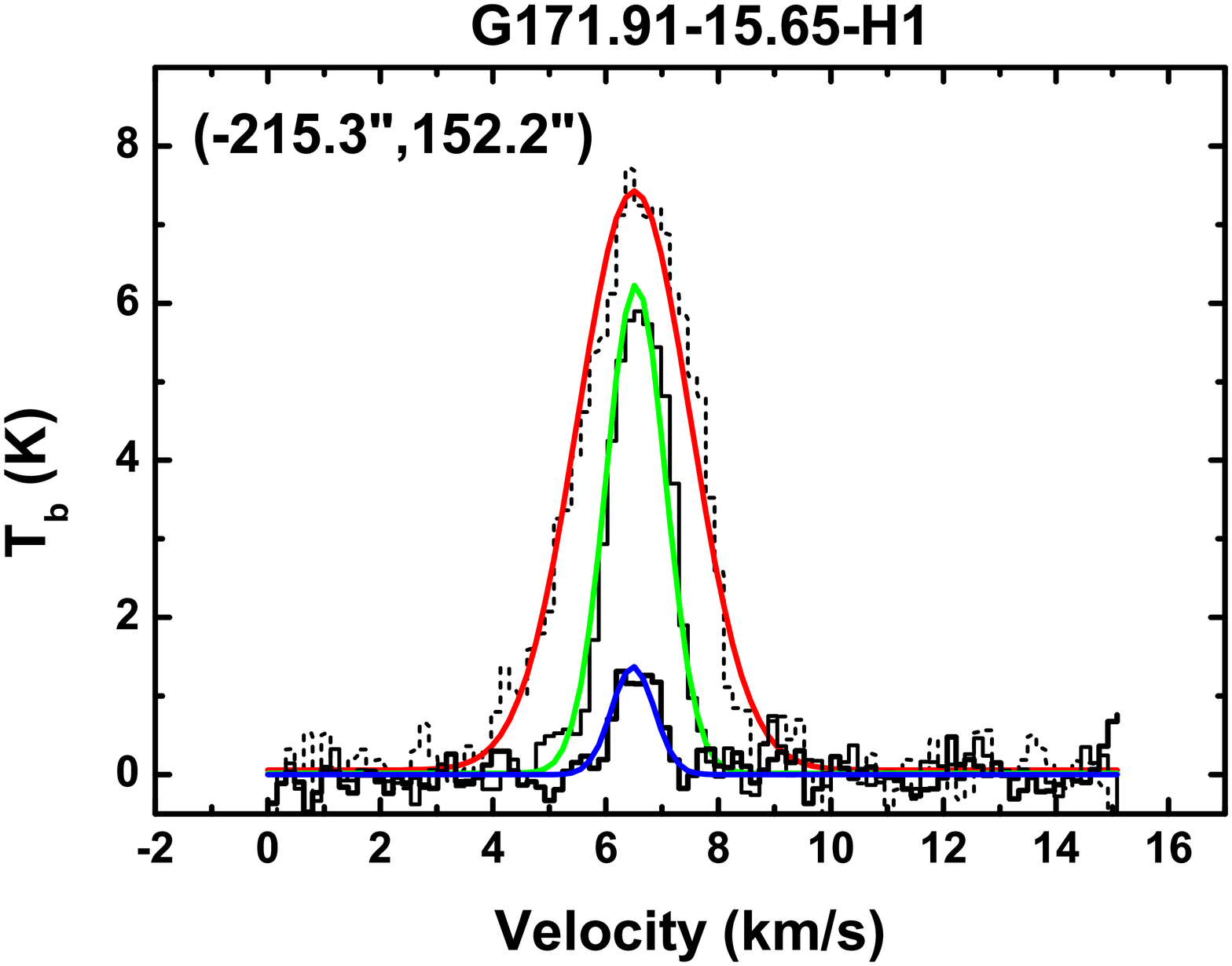}{0.3\textwidth}{(a)}
            \fig{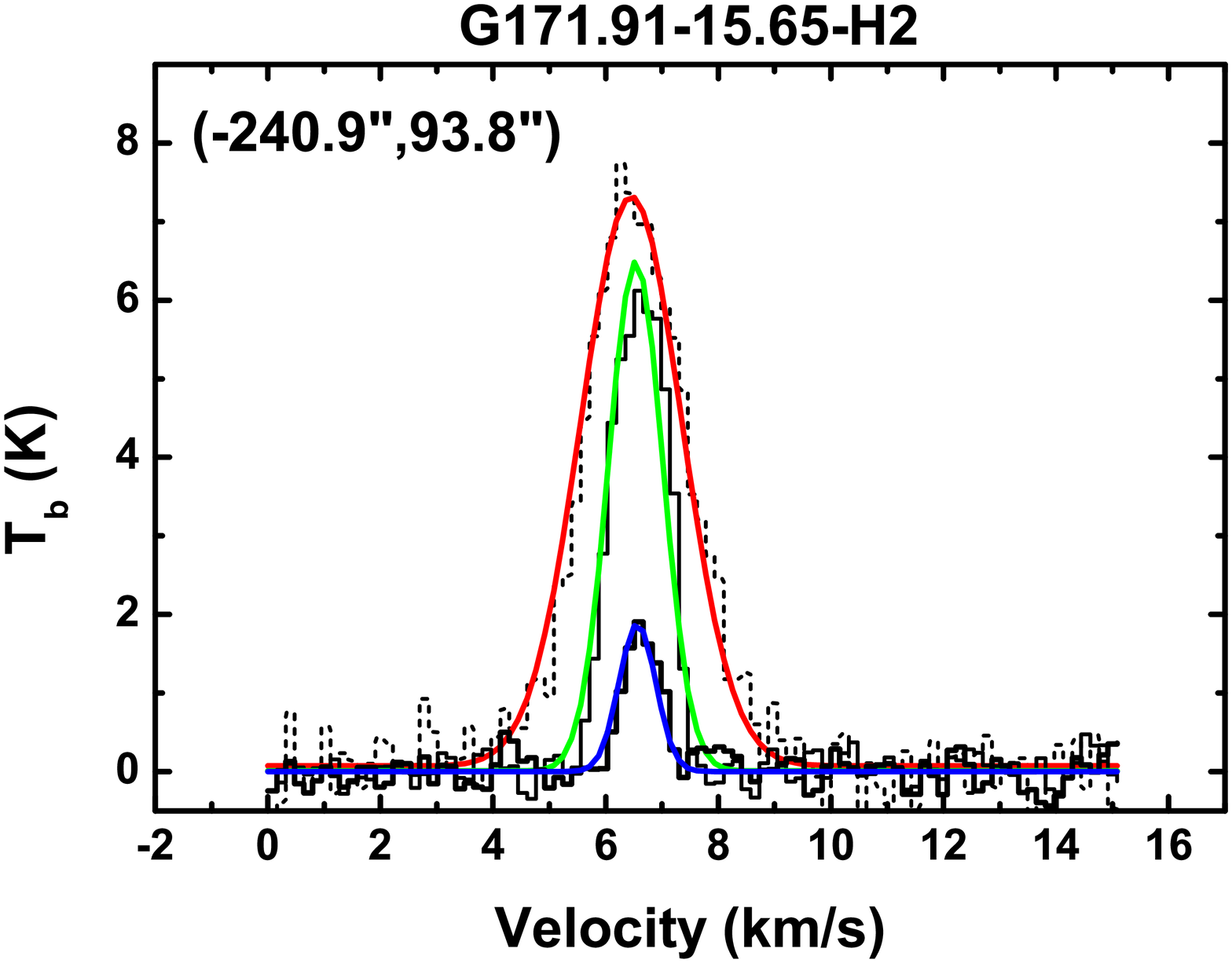}{0.3\textwidth}{(b)}
            \fig{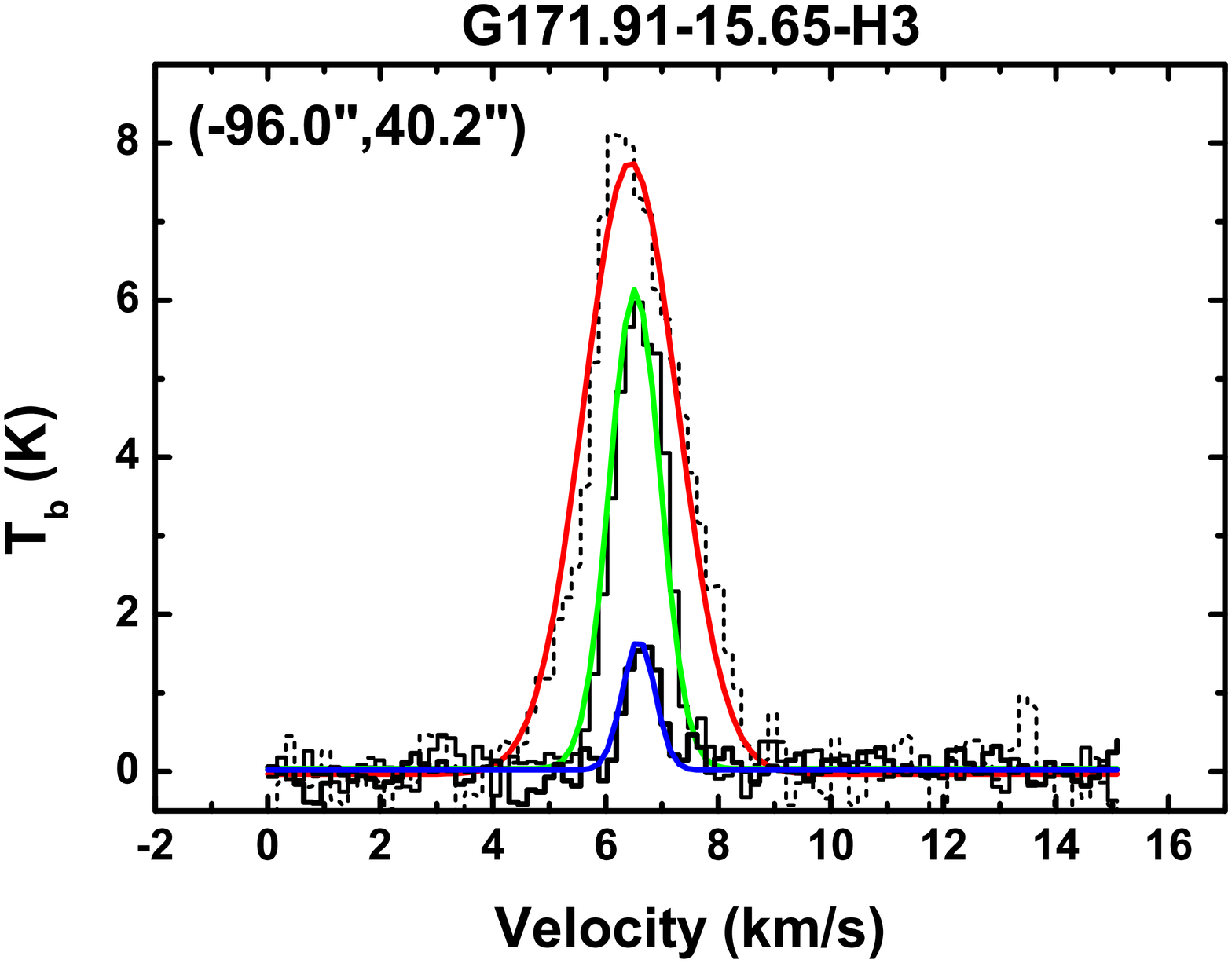}{0.3\textwidth}{(c)}
            }
\caption{Average spectra of 3 dense cores for the representative source PGCC G171.91-15.65. The images for other sources are shown in Appendix.
For each plot, the three lines of $^{12}$CO, $^{13}$CO, and C$^{18}$O are in red, green, and blue, respectively.
The offsets of dense cores are presented in the upper-left corner in unit of arc-second, and they are measured relative to coordinates of PGCCs released by \citet{Planck16}. The core names are labeled in the upper of each panel.}
\label{}
\end{figure}

\begin{figure} 
\plotone{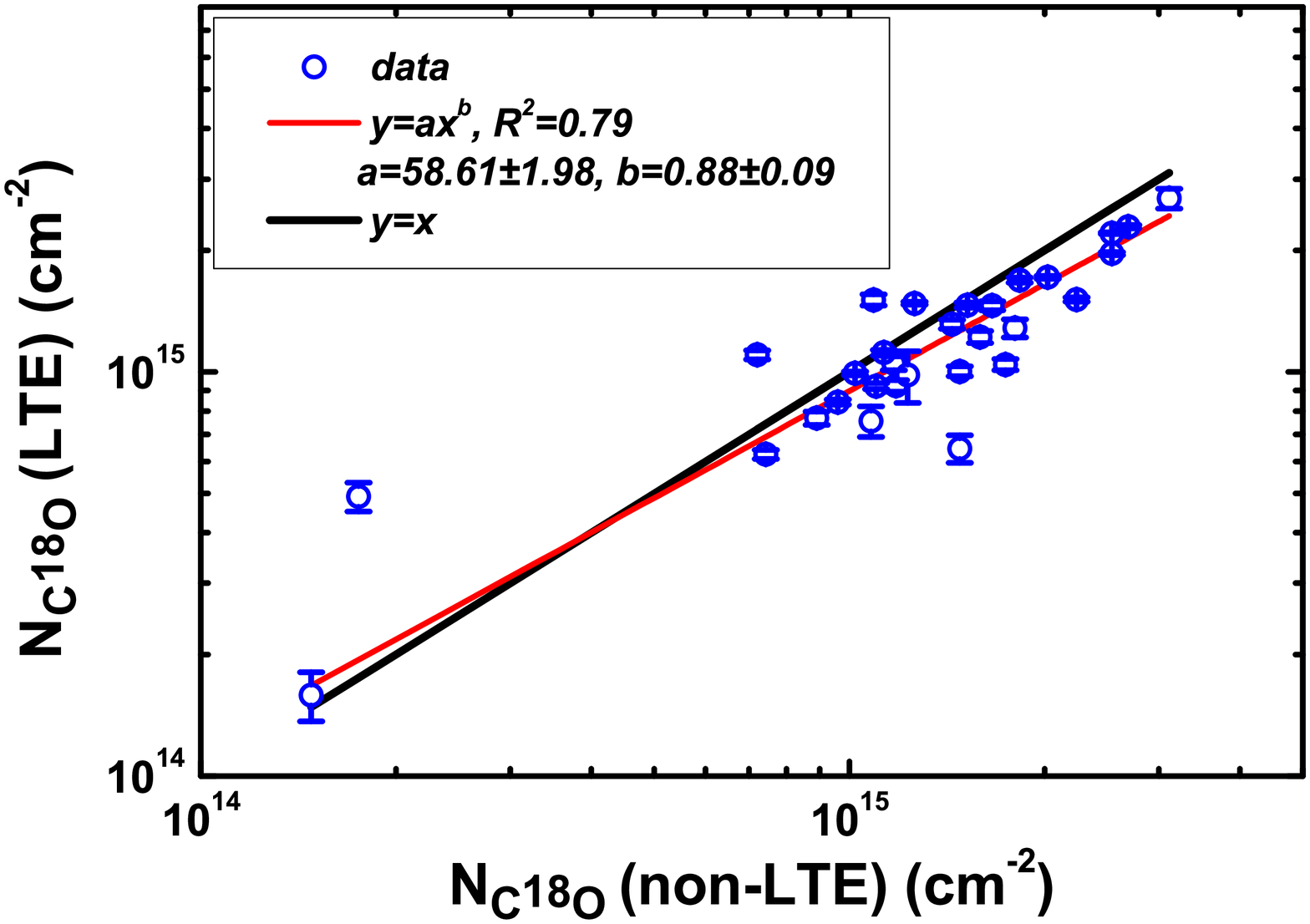}
\caption{Relationship of the LTE C$^{18}$O column density and the non-LTE C$^{18}$O column density for all of the 30 dense cores in PGCCs. The red-solid line is a power-law fit. Each blue dot represents a single dense core. The black solid line is $N_{\rm C^{18}O}$(LTE) = $N_{\rm C^{18}O}$(non-LTE). The models are presented in the upper-left corner of the figure. A strong correlation can be seen very close to $N_{\rm C^{18}O}$(LTE) = $N_{\rm C^{18}O}$(non-LTE), which indicates that the LTE C$^{18}$O column density estimates are reasonable.}
\label{}
\end{figure}

\section{DISCUSSION}

\subsection{Stabilities Of Dense Cores}
The gravitational stability is a critical factor to determine whether a dense core is gravitationally bound or not.
Assuming a Gaussian velocity distribution and an uniform density distribution, the dense core may be supported solely by random motions.
For this configuration, a virial mass can be estimated using \citep{MacLaren88}:
\begin{equation}
\it{M_{\rm vir} =  {\rm 2.10}\times 10^{\rm 2}\left(\frac{R}{\rm pc} \right)\left(\frac{\Delta V}{\rm km~s^{-1}}\right)^{\rm 2}} \rm M_{\sun},
\end{equation}
where $\Delta V$ is the FWHM of C$^{18}$O spectrum, which is derived from averaged C$^{18}$O spectra over the \emph{Herchel} dense core, and $R$ is the radius of the \emph{Herschel} dense core.
The $M_{\rm vir}$ values of 30 cores range from 3.6($\pm$0.7) M$_{\sun}$ to 74($\pm$7.6) M$_{\sun}$ (see Table 8).

We also calculated the virial parameter $\alpha$ = $M_{\rm vir}$/$M_{\rm core}$.
\citet{Kauffmann13} and \citet{Friesen16} consider $\alpha$ = 2 as a lower limit for gas motions to prevent collapse, unless the dense cores are supported by significant magnetic field or confined by external presure.
The virial parameter of the dense cores analyzed here ranges from 0.3($\pm$0.1) to 17.5($\pm$7.3).
Fourteen dense cores have virial parameters smaller than 2 and hence are
gravitationally bound and may collapse.
In Figure 7, we plot $M_{\rm core}$ against $M_{\rm vir}$ and the black dashed line indicates $M_{\rm vir}$ = 2$M_{\rm core}$.
We find that some protostellar cores and
prestellar core candidates have virial parameters larger than 2. This apparent discrepancy might be due to:

\begin{enumerate}
\item Turbulence contributions: Turbulence is only dissipated in the densest regions of clouds, while the PMO observations cannot resolve such small-scale dense regions. The size of C$^{18}$O core can be larger than the Herschel continuum core so that the C$^{18}$O line width may include significant contributions from its turbulent surroundings. For example, \cite{pat15} found that non-thermal line widths decrease substantially between the gas traced by C$^{18}$O and that traced by N$_{2}$H$^{+}$ in the dense cores of the Ophiuchus molecular cloud, indicating the dissipation of turbulence at higher densities.
\item C$^{18}$O depletion: The depletion of C$^{18}$O at the inner core region makes the C$^{18}$O(1-0) line a poor tracer at those locations. Thus, the C$^{18}$O line width is determined more by the gas outer, more turbulent regions. The presence of a central
    region affected by strong CO depletion could be the most important factor of uncertainty in the virial parameter \citep{Giannetti14}.
\end{enumerate}

Both of these factors can lead to an overestimation of the virial parameter.
C$^{18}$O lines, however, have been often used for virial analysis.
For comparison, we plot the data from \citet{Onishi96}, \citet{Tachihara00} and \citet{Zhang15} in gray stars, orange circles and black triangles respectively in Figure 7 and Figure 8. The virial masses in those works were also calculated with C$^{18}$O lines.
In contrast to dense cores in other works, the cores in L1495 have similar but relatively larger virial parameters.
We found that dense cores in far-away infrared dark clouds \citep{Zhang15,Ohashi16,Sanhueza17} have much smaller virial parameter than Taurus cores, suggesting that infrared dark clouds have much denser environments for star formation. A survey by \citet{Feher17} found 5 out of 21 PGCCs gravitationally bound based on \emph{Herschel} and CO data.

As we mentioned before, the turbulence is dissipated at denser region, C$^{18}$O is affected by turbulence from outer region, and it is also significantly affected by C$^{18}$O depletion.
Observations of this region in nitrogen-bearing tracers with comparable resolution were presented by \citet{seo15}. A total of 12 dense cores in our samples correspond to theirs. The mean value of non-thermal velocity dispersion derived from their NH$_{3}$ and our C$^{18}$O are 0.20($\pm$0.02) km s$^{-1}$ and 0.39$\pm$0.15 km s$^{-1}$, respectively. This suggests that turbulence is considerably dissipated in more centered regions. C$^{18}$O in these regions were depleted, and non-thermal velocity dispersion derived from C$^{18}$O are more contributed from diffused gas.
Together with above comparison, C$^{18}$O lines may be not good indicators of the virial parameter especially for prestellar and protostellar cores.
To clarify the situation, high-resolution observations of dense gas tracers (e.g., N$_{2}$H$^{+}$) are needed.

\citet{Kauffmann13} compiled a catalog containing 1325 virial estimates for entire molecular clouds ($\gg$ 1 pc scale), clump ($\sim$ 1 pc) and cores ($\ll$ 1 pc). They suggested an anti-correlation exists between mass and virial parameter:
\begin{equation}
\it{\alpha} = \alpha_{0}\cdot (M / \rm 10^{3} \rm ~M_{\sun})^{{\it h_{\alpha}}},
\end{equation}
with a similar slope $h_{\alpha}$, and $\alpha_{0}$ is a range of intercepts. To highlight the trend, the equation above can be rewritten as:
\begin{equation}
\it{\alpha} = \alpha_{\rm min}\cdot (M / M_{\rm max})^{h_{\alpha}},
\end{equation}
where $\alpha_{0}$ = $\alpha_{\rm min}\cdot$($M_{\rm max}$/10$^{3}$ M$_{\sun}$)$^{-h_{\alpha}}$.
We fit this power law function to our 30 samples and data from \citet{Onishi96} and \citet{Tachihara00}, the fitting results are presented in Figure 8.
Data from \citet{Zhang15} was excluded from the fitting because the number of points is small. The virial parameters from our work, \citet{Onishi96} and \citet{Tachihara00} follow above power law behavior, and with slope $h_{\alpha}$ of -0.74($\pm$0.08), -0.70($\pm$0.04), and -0.72($\pm$0.05), respectively.
This is consistent with the range 0 $<$ - $h_{\alpha}$ $<$ 1 reported in \citet{Kauffmann13}.
Similar relationships, between virial parameters and core masses are also reported by \citet{Loren89} and \citet{Bertoldi92}.
According to Figure 8, we found that these core also show similar power law behavior, indicating that these sample have similar physical properties. In Figure 7 and Figure 8, The black triangles \citep{Zhang15} have lowest virial parameters, and they show more star-forming activities such as outflows. This may indicate that our dense cores are at an evolutionary stages much earlier than cores reported by \citet{Zhang15}.

We also calculated the thermal Jeans length \citep{Jeans28} of dense cores and present in Table 8. We found that the radii of 30 dense cores are smaller than thermal Jeans length. Considering that all dense core have supersonic non-thermal motions, this may indicate that the non-thermal motions are capable of creating an effectively hydrostatic pressure opposing core collapse \citep{Mac04}.

\begin{figure}  
\plotone{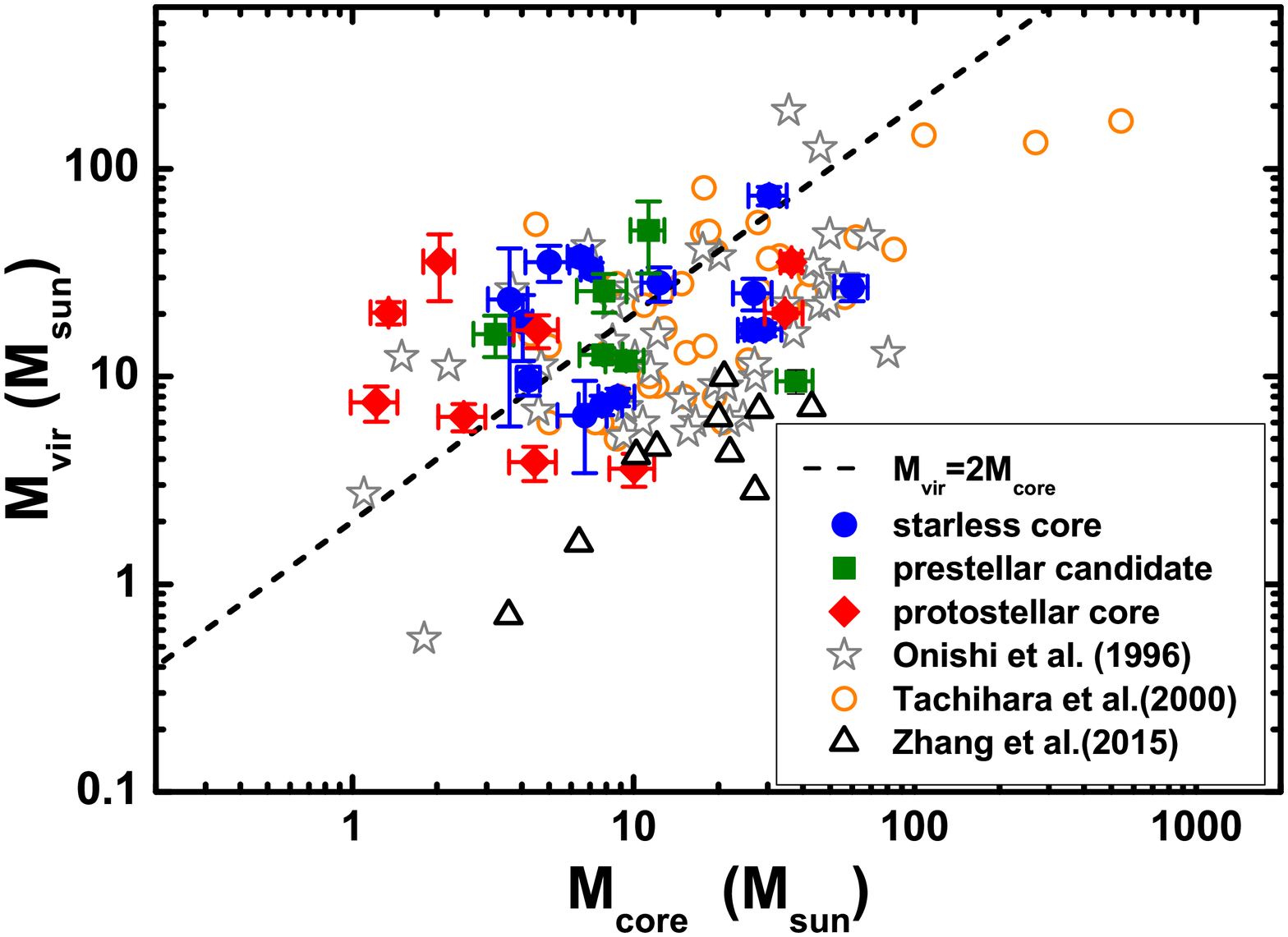}
\caption{The relationship between $M_{\rm core}$ and $M_{\rm vir}$ of dense cores. The red diamonds, green squares and blue dots represent the protostellar cores, prestellar core candidates and starless cores, respectively. The gray stars, orange circles, and black triangles are data from \citet{Onishi96}, \citet{Tachihara00}, and \citet{Zhang15}, respectively. The black dashed line indicates $M_{\rm vir}$ = 2$M_{\rm core}$. The virial parameter $\alpha$ = 2 can be regarded as a lower limit for gas motions to prevent collapse.}
\label{}
\end{figure}

\begin{figure}  
\plotone{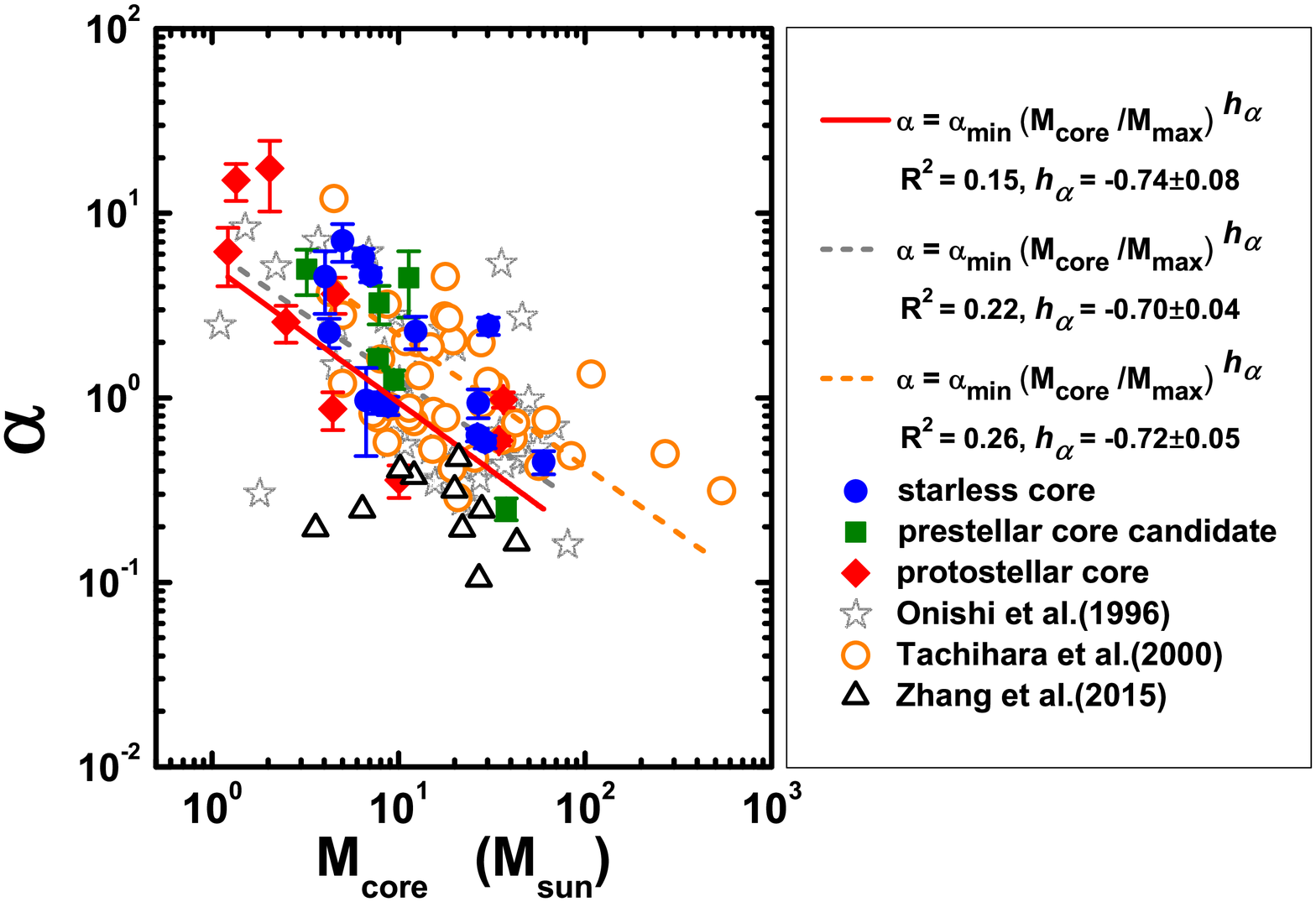}
\caption{The relationship between $M_{\rm core}$ and $\alpha$. The red diamonds, green squares and blue dots represent the protostellar cores, prestellar core candidates and starless coress, respectively.  The gray stars, orange circles, and black triangles are data from \citet{Onishi96}, \citet{Tachihara00}, and \citet{Zhang15} respectively. The red solid, gray and orange dashed lines indicate power law fitting of $\alpha$ = $\alpha_{\rm min}\cdot$($M_{\rm core}$/$M_{\rm max}$)$^{h_{\alpha}}$ for L1495 samples, gray stars, and orange circles, respectively.}
\label{}
\end{figure}

\subsection{Column Density Profile}
In this section, we study the density profiles of the dense cores. We averaged the \emph{Herschel} column density data in
concentric elliptical bins to get a column density profile, and then fitted the Spherical Geometry Model of \citet{Dapp09}.
The underlying model resembles the Bonnor-Ebert model in that it features a flat central region leading into a power-law decline $\propto$ r$^{-2}$ in density, and a well-defined outer radius. The model can overcome the Bonnor-Ebert profile in several aspects. However, this model does not assume that the cloud is in equilibrium, and can instead make qualitative statements about its dynamical state (expansion, equilibrium, collapse) using the size of the flat region as a proxy \citep{Dapp09}. Therefore, it is more suitable to study the properties of dense cores in different dynamical states.

The basic model is:
\begin{equation}
n(r)=\begin{cases}
 n_{\rm c} a^2 / (r^2 + a^2) & r \leq R_{\rm t} \\
0   &   r > R_{\rm t},
\end{cases}
\end{equation}
which is characterized by a central volume density $n_{\rm c}$ and a truncated radius $R_{\rm t}$ (In order to avoid confusion with the core radius from Gaussian fits, we use $R_{\rm t}$ to represent the truncated radius of cores derived from density profiles). The parameter \emph{a} represents the size of the inner flat region of a dense core. The column density $N_{x}$ can be derived by integrating the volume density along the line of sight through the sphere. Introducing the parameter $c \equiv R_{\rm t}/a$, $N_{\rm c} = 2 a n_{\rm c}\arctan(c)$. Hence, the model can be re-written as:
\begin{align}
  N_{x} = \frac{N_{\rm c}}{\sqrt{1+(x/a)^2}} \times \left [ \arctan(\sqrt{\frac{c^2-(x/a)^2}{1+(x/a)^2}})/\arctan(c)  \right ].
\end{align}
where $N_{\rm c}$, $R_{\rm t}$, and \emph{a} are free parameters. Due to their small sizes, we can not obtain average column density distributions for G170.26-16.02-H1 and G171.91-16.15-H1. Therefore, we only fit \emph{Herschel} column density profiles to the other 28 cores in this section.
$N_{\rm c}$, $R_{\rm t}$, \emph{a}, and $n_{\rm c}$ are shown in Table 9, and statistics of these parameters are summarized in Table 10.

The beam size is a critical factor that can affect the column density profile.
For example, the resolution of the column density maps from \emph{Herschel} data sets a lower limit for the size of the flat region \emph{a} and the core truncated radius $R_{\rm t}$.
To remove beam effect, we use a quadratic rule, $a_{\rm d}$ = $(a^{2}-\theta_{beam}^{2})^{1/2}$, to compute the deconvolved size of the flat region $a_{\rm d}$.
A similar method was also used to calculate the deconvolved core radius $R_{\rm d}$ ($a_{\rm d}$ and $R_{\rm d}$ are listed in Table 9).
Since measured flat region of G171.91-15.65-H2 is smaller than the beam size, we cannot compute its $a_{\rm d}$, and that core was also excluded from further analyses. Therefore, we present the column density profiles of G171.91-15.65-H3 in Figure 9. The profiles for other dense cores are shown in Appendix.

The upper, middle, and bottom panels of Figure 10 present the average density profiles of starless cores, prestellar core candidates, and protostellar cores, respectively.
We normalized each respective core type's column density profile before averaging.
The truncated radius $R_{\rm t}$, decreases as dense cores evolve from starless to protostellar.
The size of the central flat region (\emph{a}) is significantly larger in the starless cores (the mean value of \emph{a} is 0.48 pc for starless cores), but comparable between the prestellar candidates and protostellar cores (both the value of \emph{a} for prestellar candidates and protostellar cores are 0.26 pc). This indicates that the starless cores are less peaked than prestellar core candidates or protostellar cores.

Figure 11 presents $R_{\rm d}$, $a_{\rm d}$, and $n_{\rm c}$. The upper panel of Figure 11 presents the results from fits to the column density profile for each core. The red, green, and blue dots are protostellar cores, prestellar core candidates, and starless cores, respectively. The bottom panel shows the values from fits to averaged protostellar cores, prestellar core candidates, and starless cores. According to the
statistics of the fitting results (see Table 10), the denser the cores are, the smaller the values of $R_{\rm d}$ and \emph{a} are. In other words, the dense cores will show more steeper density structures when they evolve from starless cores to prestellar or protostellar cores.
\begin{deluxetable}{cccccccccccccccccccccccccccccccccc} 
\tabletypesize{\scriptsize} \tablecolumns{11}
\tablewidth{0pc} \setlength{\tabcolsep}{0.05in}
\tablecaption{The parameters derived from column density profiles}
\tablehead{
 Name &$N_{\rm c}$               &$R_{\rm t}$\tablenotemark{a}        &\emph{a}       &$n_{\rm c}$\tablenotemark{b}     &$a_{\rm d}$\tablenotemark{c}  &$R_{\rm d}$\tablenotemark{d}            \\
      &(10$^{22}$ cm$^{-2}$) &(pc)     &(pc)    &(10$^{4}$ cm$^{-3}$)    &(pc) &(pc) \\
}
\startdata
G166.99-15.34-H1  &0.66($\pm$0.01)   &0.12($\pm$ 0.04)  &0.028($\pm$0.002)   &2.88($\pm$0.38)    &0.025($\pm$0.002) &0.12($\pm$0.04)        \\
G167.23-15.32-H1  &1.19($\pm$0.01)   &0.10($\pm$ 0.01)  &0.045($\pm$0.001)   &3.77($\pm$0.12)    &0.043($\pm$0.001) &0.10($\pm$0.01)        \\
G168.00-15.69-H1  &0.91($\pm$0.01)   &0.28($\pm$0.04)   &0.089($\pm$0.005)   &1.31($\pm$0.11)    &0.089($\pm$0.005) &0.28($\pm$0.04)        \\
G168.13-16.39-H1  &2.68($\pm$0.01)   &0.38($\pm$0.04)   &0.038($\pm$0.002)   &7.79($\pm$0.45)    &0.036($\pm$0.002) &0.40($\pm$0.04)        \\
G168.13-16.39-H2  &1.62($\pm$0.01)   &0.25($\pm$0.03)   &0.043($\pm$0.006)   &4.35($\pm$0.64)    &0.042($\pm$0.006) &0.25($\pm$0.03)        \\
G168.72-15.48-H1  &2.10($\pm$0.01)   &0.93($\pm$0.67)   &0.048($\pm$0.017)   &4.65($\pm$1.75)    &0.047($\pm$0.017) &0.92($\pm$0.67)        \\
G168.72-15.48-H2  &4.51($\pm$0.02)   &0.37($\pm$0.25)   &0.028($\pm$0.001)   &17.74($\pm$1.30)   &0.025($\pm$0.001) &0.37($\pm$0.25)        \\
G169.43-16.17-H1  &1.51($\pm$0.01)   &0.19($\pm$0.04)   &0.083($\pm$0.007)   &2.56($\pm$0.35)    &0.082($\pm$0.007) &0.19($\pm$0.04)        \\
G169.43-16.17-H2  &1.41($\pm$0.01)   &0.09($\pm$0.04)   &0.039($\pm$0.017)   &5.02($\pm$2.23)    &0.037($\pm$0.018) &0.09($\pm$0.04)        \\
G169.76-16.15-H1  &1.20($\pm$0.01)   &0.88($\pm$0.44)   &0.048($\pm$0.006)   &2.65($\pm$0.30)    &0.047($\pm$0.005) &0.88($\pm$0.44)        \\
G169.76-16.15-H2  &1.13($\pm$0.01)   &0.99($\pm$0.36)   &0.053($\pm$0.005)   &2.26($\pm$0.24)    &0.052($\pm$0.005) &0.99($\pm$0.36)        \\
G169.76-16.15-H3  &1.80($\pm$0.01)   &0.29($\pm$0.11)   &0.029($\pm$0.001)   &6.95($\pm$0.43)    &0.026($\pm$0.001) &0.29($\pm$0.11)        \\
G169.76-16.15-H4  &1.73($\pm$0.01)   &0.29($\pm$0.23)   &0.029($\pm$0.002)   &6.50($\pm$0.72)    &0.027($\pm$0.002) &0.29($\pm$0.23)        \\
G169.76-16.15-H5  &2.18($\pm$0.03)   &0.10($\pm$0.05)   &0.018($\pm$0.001)   &14.06($\pm$1.93)   &0.014($\pm$0.002) &0.10($\pm$0.05)        \\
G170.00-16.14-H1  &1.86($\pm$0.02)   &0.09($\pm$0.02)   &0.026($\pm$0.002)   &9.11($\pm$1.03)    &0.023($\pm$0.002) &0.09($\pm$0.02)        \\
G170.00-16.14-H2  &2.36($\pm$0.03)   &0.06($\pm$0.01)   &0.022($\pm$0.002)   &14.19($\pm$1.35)   &0.019($\pm$0.002) &0.06($\pm$0.01)        \\
G170.13-16.06-H1  &1.61($\pm$0.04)   &0.13($\pm$0.01)   &0.085($\pm$0.012)   &3.11($\pm$0.44)    &0.084($\pm$0.012) &0.13($\pm$0.01)        \\
G170.26-16.02-H2  &2.30($\pm$0.01)   &0.12($\pm$0.01)   &0.027($\pm$0.001)   &10.41($\pm$0.33)   &0.024($\pm$0.001) &0.12($\pm$0.01)        \\
G170.83-15.90-H1  &0.72($\pm$0.01)   &0.13($\pm$0.03)   &0.049($\pm$0.004)   &1.95($\pm$0.26)    &0.048($\pm$0.004) &0.13($\pm$0.03)        \\
G170.83-15.90-H2  &0.96($\pm$0.01)   &0.18($\pm$0.09)   &0.032($\pm$0.002)   &3.47($\pm$0.38)    &0.030($\pm$0.002) &0.18($\pm$0.09)        \\
G170.99-15.81-H1  &1.36($\pm$0.01)   &0.26($\pm$0.05)   &0.050($\pm$0.002)   &3.16($\pm$0.22)    &0.049($\pm$0.002) &0.26($\pm$0.05)        \\
G171.49-14.90-H1  &3.79($\pm$0.03)   &0.11($\pm$0.01)   &0.025($\pm$0.001)   &18.11($\pm$0.56)   &0.022($\pm$0.001) &0.11($\pm$0.01)        \\
G171.80-15.32-H1  &2.61($\pm$0.04)   &0.10($\pm$0.01)   &0.031($\pm$0.002)   &10.65($\pm$0.73)   &0.029($\pm$0.002) &0.10($\pm$0.01)        \\
G171.80-15.32-H2  &1.37($\pm$0.01)   &0.17($\pm$0.08)   &0.027($\pm$0.002)   &5.96($\pm$0.72)    &0.024($\pm$0.002) &0.16($\pm$0.08)        \\
G171.80-15.32-H3  &1.91($\pm$0.01)   &0.23($\pm$0.03)   &0.030($\pm$0.001)   &7.13($\pm$0.28)    &0.028($\pm$0.001) &0.23($\pm$0.03)        \\
G171.91-15.65-H2$\dagger$  &2.43($\pm$0.06)   &0.11($\pm$0.08)   &0.010($\pm$0.001)   &27.29($\pm$4.08)   &...      &0.11($\pm$0.08)        \\
G171.91-15.65-H3  &1.00($\pm$0.01)   &0.90($\pm$0.27)   &0.043($\pm$0.002)   &2.46($\pm$0.13)    &0.041($\pm$0.002) &0.90($\pm$0.27)        \\
G172.06-15.21-H1  &1.51($\pm$0.03)   &0.27($\pm$0.09)   &0.025($\pm$0.002)   &6.76($\pm$0.67)    &0.021($\pm$0.002) &0.27($\pm$0.08)        \\
\enddata
\tablenotetext{a}{The truncated radius are derived from fitting to column density profile.}
\tablenotetext{b}{The central volume density calculated by: $n_{\rm c} = N_{\rm c} / 2 a \arctan(c)$, $c \equiv R/a$}
\tablenotetext{c}{$a_{\rm d}$ represents the flat region which were removed beam effect by using $a_{\rm d}$ = $(a^{2}-beam^{2})^{1/2}$.}
\tablenotetext{d}{$R_{\rm d}$ represents the flat region which were removed beam effect by using $R_{\rm d}$ = $(R^{2}-beam^{2})^{1/2}$.}
\tablenotetext{\dagger}{The flat region \emph{a} of G171.91-15.65-core2 is smaller than beam size (0.012 pc), we can not calculate the $a_{\rm d}$ for this cores.}
\tablecomments{All of these parameters are come from fitting to column density profile. The column density profiles are derived from \emph{Herschel} column density map. Due to small size of G170.26-16.02-H1 and G171.91-16.15-H1, we can not obtain their average column density distributions. Thus, we only present the parameters of 28 dense cores in this table.}
\end{deluxetable}

\begin{splitdeluxetable*}{ccccccccccccBccccccccccccccc} 
\tabletypesize{\tiny}  \tablecolumns{27}
\tablewidth{0pc} \setlength{\tabcolsep}{0.04in}
\tablecaption{Statistics of parameters} \tablehead{
Classification & &$N_{H_{2}}$(Herschel) &$N_{H_{2}}$(PMO) &$T_{\rm d}$ &$V_{\rm lsr}$ &$\sigma_{\rm th}$  &$\sigma_{\rm NT}$ ($^{13}$CO) &$\sigma_{\rm 3D}$ &$\sigma_{\rm NT}$ (C$^{18}$O) &Mach number &\emph{R(Herschel)} &Classification & &$M_{\rm core}$ &$M_{\rm vir}$ &$\alpha$ &Jeans length &$n_{\rm H_{2}}$  &$N_{\rm C^{18}O}$ &C$^{18}$O abundance  &$f_{\rm D}$ &$a_{\rm d}$ &$N_{\rm c}$ &$n_{\rm c}$ \\
     & &(10$^{22}$ cm$^{-2}$) &(10$^{22}$ cm$^{-2}$) &(K) &(km s$^{-1}$)  &(km s$^{-1}$) &(km s$^{-1}$) &(km s$^{-1}$) &(km s$^{-1}$) & &(pc) & & &(M$_\sun$) &(M$_\sun$) & &(pc) &(10$^{4}$ cm$^{-3}$) &(10$^{14}$ cm$^{-2}$) &(10$^{-7}$)  & &(pc) &(10$^{22}$ cm$^{-2}$) &(10$^{4}$ cm$^{-3}$) }
\startdata
{              } &Mean   &1.10 &0.40 &12.1 &6.57 &0.19 &0.51 &0.94 &0.38 &2.08  &0.12  &{         } &Mean   &13.76 &21.43 &3.49  &0.216 &2.88  &12.86 &1.64 &1.56 &0.038  &1.78 &6.63  \\
{all cores     } &Max    &2.46 &0.68 &13.8 &7.78 &0.21 &0.73 &1.32 &0.71 &3.83  &0.31  &{all cores} &Max    &59.83 &74.28 &17.46 &0.440 &7.87  &22.89 &2.87 &6.59 &0.089  &4.51 &18.11 \\
{              } &Min    &0.36 &0.13 &11.1 &5.37 &0.16 &0.36 &0.70 &0.19 &1.00  &0.044 &{         } &Min    &1.21  &3.60  &0.25  &0.111 &0.57  &1.58  &0.66 &1.00 &0.014 &0.66 &1.31  \\
{              } &Median &1.05 &0.38 &12.0 &6.53 &0.20 &0.51 &0.95 &0.36 &1.94  &0.11  &{         } &Median &7.75  &17.55 &2.28  &0.206 &2.33  &12.52 &1.65 &2.24 &0.030  &1.61 &5.01  \\
\hline
{              } &Mean   &0.86 &0.38 &12.1 &6.48 &0.19 &0.50 &0.94 &0.36 &1.99 &0.15  &{               } &Mean   &15.90 &24.46 &2.73  &0.268 &1.48  &13.27 &1.57  &1.86 &0.048 &1.27 &3.50 \\
{starless cores} &Max    &1.40 &0.68 &13.1 &7.78 &0.21 &0.73 &1.32 &0.56 &3.04 &0.31  &{ starless cores} &Max    &59.83 &74.28 &7.09  &0.440 &2.53  &22.89 &2.87  &6.59 &0.089 &2.10 &6.76 \\
{              } &Min    &0.36 &0.13 &11.5 &5.37 &0.16 &0.36 &0.71 &0.22 &1.07 &0.096 &{               } &Min    &3.62  &6.48  &0.45   &0.189 &0.57   &1.58  &0.44  &1.00 &0.021 &0.66 &1.31 \\
{              } &Median &0.86 &0.38 &12.1 &6.49 &0.18 &0.49 &0.91 &0.34 &1.82 &0.12  &{               } &Median &7.73  &23.49 &2.27   &0.240 &1.57   &14.58 &1.61  &1.74 &0.047 &1.20 &3.11 \\
\hline
{                }  &Mean   &1.47 &0.43 &11.6 &6.76 &0.19 &0.54 &1.00 &0.42 &2.21 &0.099  &{                } &Mean   &12.89 &21.04 &2.64 &0.158 &3.96  &12.03 &0.81 &3.66 &0.026 &2.50 &10.33 \\
{prestellar candidates}  &Max    &2.46 &0.68 &13.2 &7.19 &0.21 &0.67 &1.21 &0.62 &3.41 &0.14   &{prestellar candidates} &Max    &37.79 &50.51 &4.97 &0.202 &5.01  &22.32 &0.99 &4.70 &0.039 &4.51 &17.74 \\
{                }  &Min    &1.05 &0.35 &11.1 &5.89 &0.18 &0.43 &0.81 &0.23 &1.11 &0.061  &{                } &Min    &3.22  &9.49  &0.25 &0.128 &2.53  &0.76  &0.61 &2.91 &0.023 &1.80 &6.94  \\
{                }  &Median &1.29 &0.38 &11.1 &6.97 &0.19 &0.52 &0.96 &0.41 &2.26 &0.098  &{                } &Median &8.63  &14.33 &2.45 &0.158 &4.43  &10.82 &0.84 &3.41 &0.025 &1.10 &9.76  \\
\hline
{                  } &Mean   &1.22 &0.39 &12.4 &6.56 &0.19 &0.47 &0.87 &0.40 &2.14 &0.090  &{                  } &Mean   &10.79 &16.64 &5.30  &0.167 &4.49  &12.73 &1.12 &3.07 &0.026 &2.34 &10.74 \\
{protostellar cores} &Max    &2.12 &0.47 &13.8 &7.10 &0.19 &0.64 &1.16 &0.71 &3.83 &0.22   &{protostellar cores} &Max    &36.43 &35.66 &17.46 &0.244 &7.87  &19.63 &1.64 &5.97 &0.042 &3.79 &18.11 \\
{                  } &Min    &0.78 &0.25 &11.3 &5.97 &0.18 &0.36 &0.70 &0.19 &1.00 &0.044  &{                  } &Min    &1.21  &3.60  &0.36  &0.111 &1.22  &0.63  &0.48 &1.75 &0.014&1.37 &4.35  \\
{                  } &Median &1.13 &0.39 &12.2 &6.49 &0.19 &0.43 &0.84 &0.37 &1.99 &0.065  &{                  } &Median &4.50  &16.68 &2.32  &0.162 &4.12  &12.82 &1.28 &2.30 &0.023 &2.27 &10.91 \\
\hline
\enddata
\tablecomments{The statistics of main parameters for our analysis results. For all parameters, we represent the mean, maximum, minimum and median values for all cores, starless cores, prestellar candidates, and protostellar cores, respectively. However, we have not included the SCUBA-2 data in this statistic, because only a part of dense cores have been detected by SCUBA-2.}
\end{splitdeluxetable*}

\begin{figure} 
\plotone{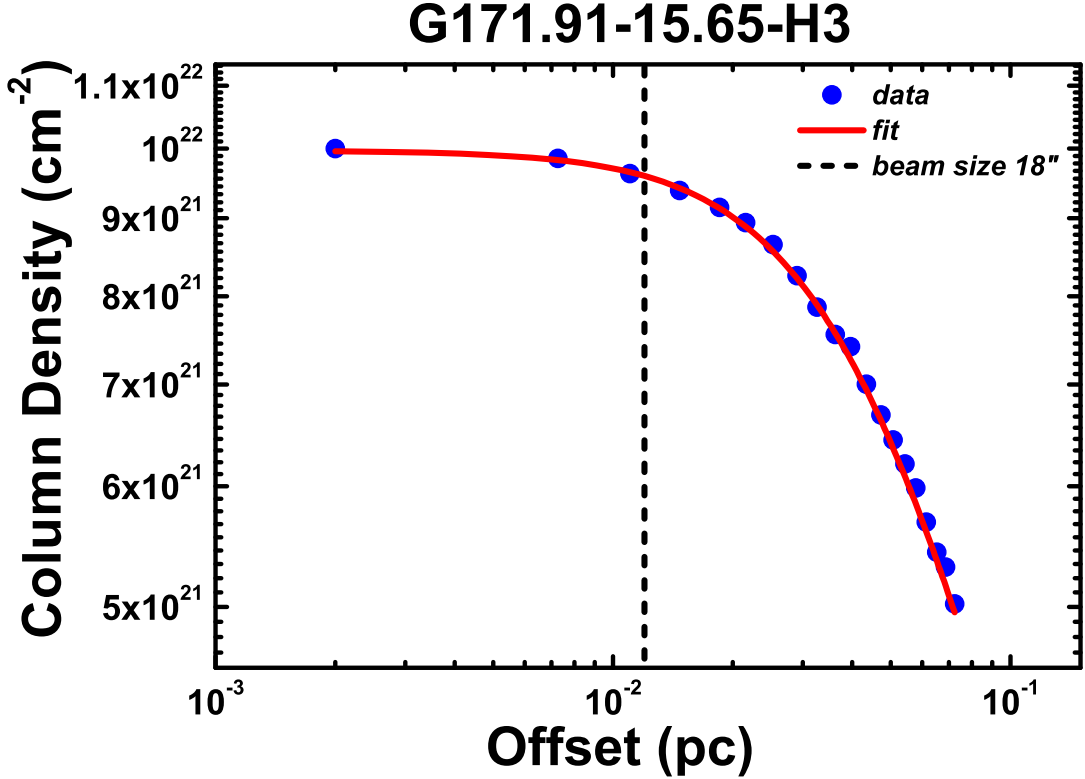}
\caption{The column density profiles of dense core for G171.91-15.65-H3, red solid line represents the corresponding fit. Fitting results of $N_{\rm c}$, R and \emph{a} are listed in Table 9. The flat central region and power-law decline can be seen in figure. Due to its small size, we can not obtain average column density distributions accurately for G170.26-16.02-H1 and G171.91-16.15-H1, and the flat region \emph{a} of G171.91-15.65-H1 is smaller than beam size. Therefore, we only present the column density profile for G171.91-15.65-H3 in this section. The images for other sources are shown in Appendix.}
\label{}
\end{figure}

\begin{figure*} 
\gridline{\fig{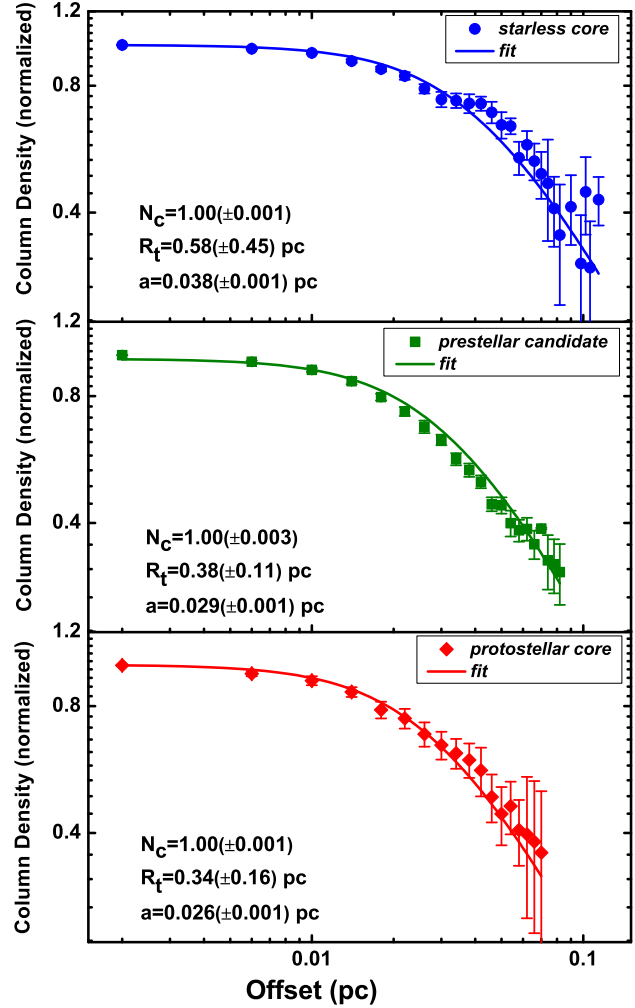}{0.5\textwidth}{}
          }
\caption{The average column density profile of starless cores (upper panel), prestellar core candidates (middle panel) and protostellar cores (bottom panel), respectively. Fitting results of $N_{\rm c}$, $R_{\rm t}$, and \emph{a} are shown in each panel.}
\label{}
\end{figure*}
\begin{figure} 
\plotone{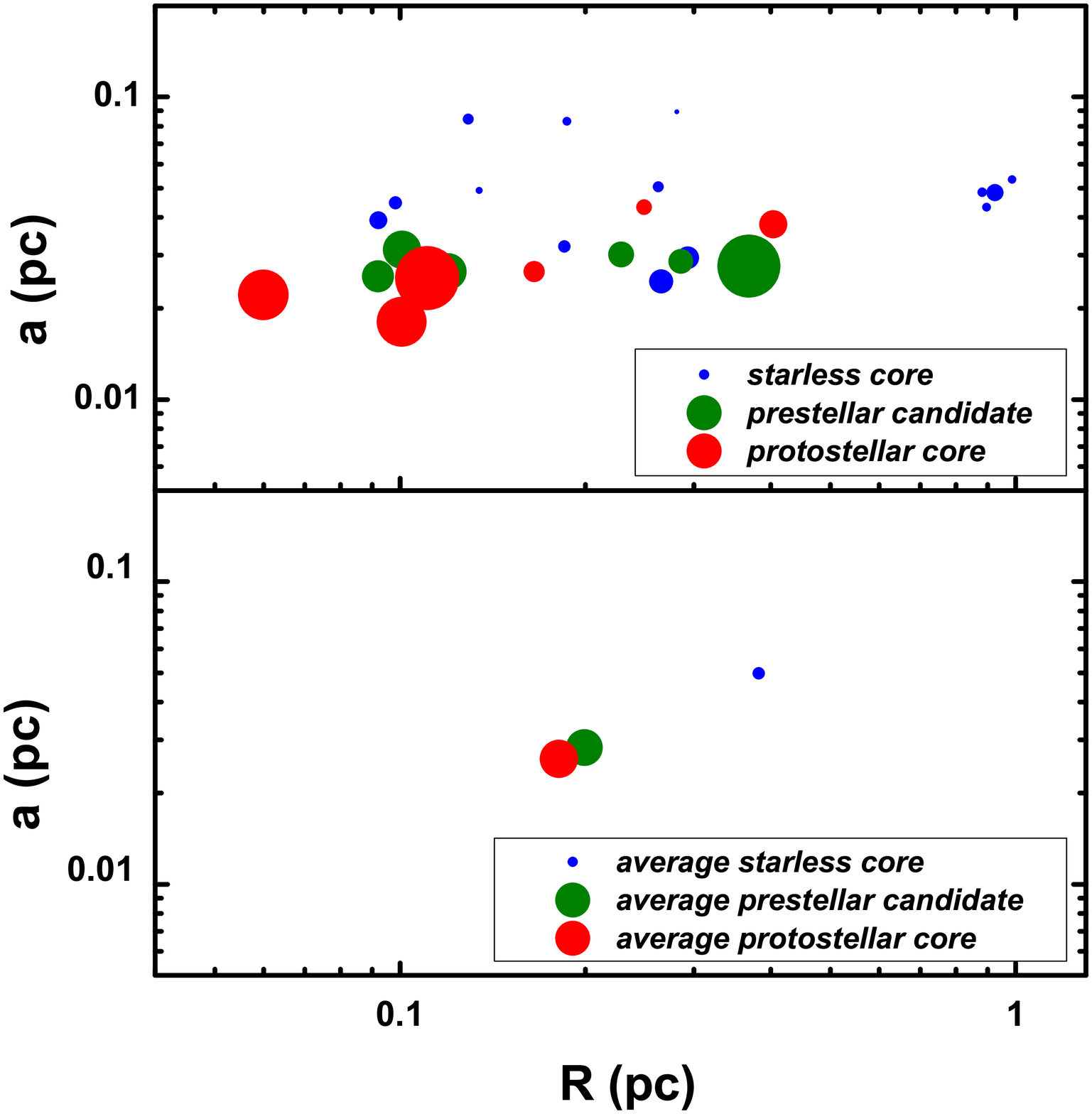}
\caption{\emph{Top}: the relationships between $R_{\rm d}$, $a_{\rm d}$, and $n_{\rm c}$ for L1495 dense cores. The \emph{x} axis represents the radius $R_{\rm d}$, the \emph{y} axis represents the flat region $a_{\rm d}$. The relative central volume densities $n_{\rm c}$ range from 1.3$\times$10$^{4}$ to 1.8$\times$10$^{5}$ cm$^{-3}$ are shown by the sizes of the data points. \emph{Bottom}: The relationships between $R_{\rm d}$, $a_{\rm d}$, and $n_{\rm c}$ of averaged protostellar cores, prestellar core candidates and starless cores. In both panels, the red, green, and blue dots represent protostellar cores, prestellar core candidates, and starless cores, respectively.}
\label{}
\end{figure}

\subsection{CO Depletion}
Gaseous CO molecules can freeze out onto grain surfaces in cold and
dense regions, and the CO molecules can return to the gas phase as
temperature rises due to heating from protostars \citep{Bergin97,Charnley97,Zhang17}.

Since C$^{18}$O emission is generally optically thin, it is a more reliable tracer for CO depletion than $^{12}$CO or $^{13}$CO. We calculate the core-averaged C$^{18}$O abundance of each dense core as follows:
\begin{equation}
X = \frac{N_{\rm C^{18}O}}{N_{\rm H_{2}}(Herschel)},
\end{equation}
where $N_{\rm H_{2}}$ is the core's \emph{Herschel}-derived column density of H$_{\rm 2}$ (listed in Table 3), and $N_{\rm C^{18}O}$ is the C$^{18}$O column density.
It should be noted that the G166.99-15.34-H1 and G167.23-15.32-H1 are located outside of the densest parts of the filament, and with exceptionally low C$^{18}$O abundances.
The low C$^{18}$O abundance in these two cores may be because that they are less shielded from interstellar UV radiation and the CO molecules inside them are more easily dissociated.
Thus, C$^{18}$O abundance of these two dense cores are very low, and they were excluded from fitting.
In Figure 12, the red diamonds, green squares and blue dots represent the protostellar cores, prestellar core candidates and starless cores, respectively.
C$^{18}$O abundance is clearly anti-correlated with the column density, also indicating that the CO depletion becomes more prevalent in denser cores.

In this work, we simply define a relative CO depletion factor ($f_{\rm D}$) of dense cores as follows:
\begin{equation}
{f_{\rm D} = \frac{X_{\rm max}}{X_{\rm core}}},
\end{equation}
where $X_{\rm max}$ is the maximum C$^{18}$O abundance in all the L1495 dense cores, and $X_{\rm core}$ is the C$^{18}$O abundance of dense core.
The CO depletion factors ($f_{\rm D}$) range from 1 to 6.6($\pm$1.8), and for each core these parameters are listed in Table 7. Please note that the CO depletion factors derived here should be treated as lower limits because we used core-averaged values. The CO depletion could be even more severe toward the core center.

Figure 13 presents $f_{\rm D}$ and the central volume densities ($n_{\rm c}$) of the L1495 dense cores.
There is a clear positive correlation between $f_{\rm D}$ and $n_{\rm c}$, indicating that CO depletion is more significant in denser cores.
The starless cores are located at bottom-left in Figure 13 and thus these cores have smallest CO depletion factors. CO depletion in prestellar candidates and protostellar cores is more significant than starless cores.
The mean depletion factors of starless cores, prestellar core candidates, and protostellar cores are 1.9($\pm$0.7), 3.7($\pm$0.7), and 3.1($\pm$1.5), respectively.
The prestellar candidates and protostellar cores are mixed with each other in the Figure 12 and Figure 13, indicating that the protostellar cores in L1495 are still at the earliest phases of star formation and thus have as a high degree of CO depletion as prestellar core candidates.

\begin{figure}[!ht] 
\plotone{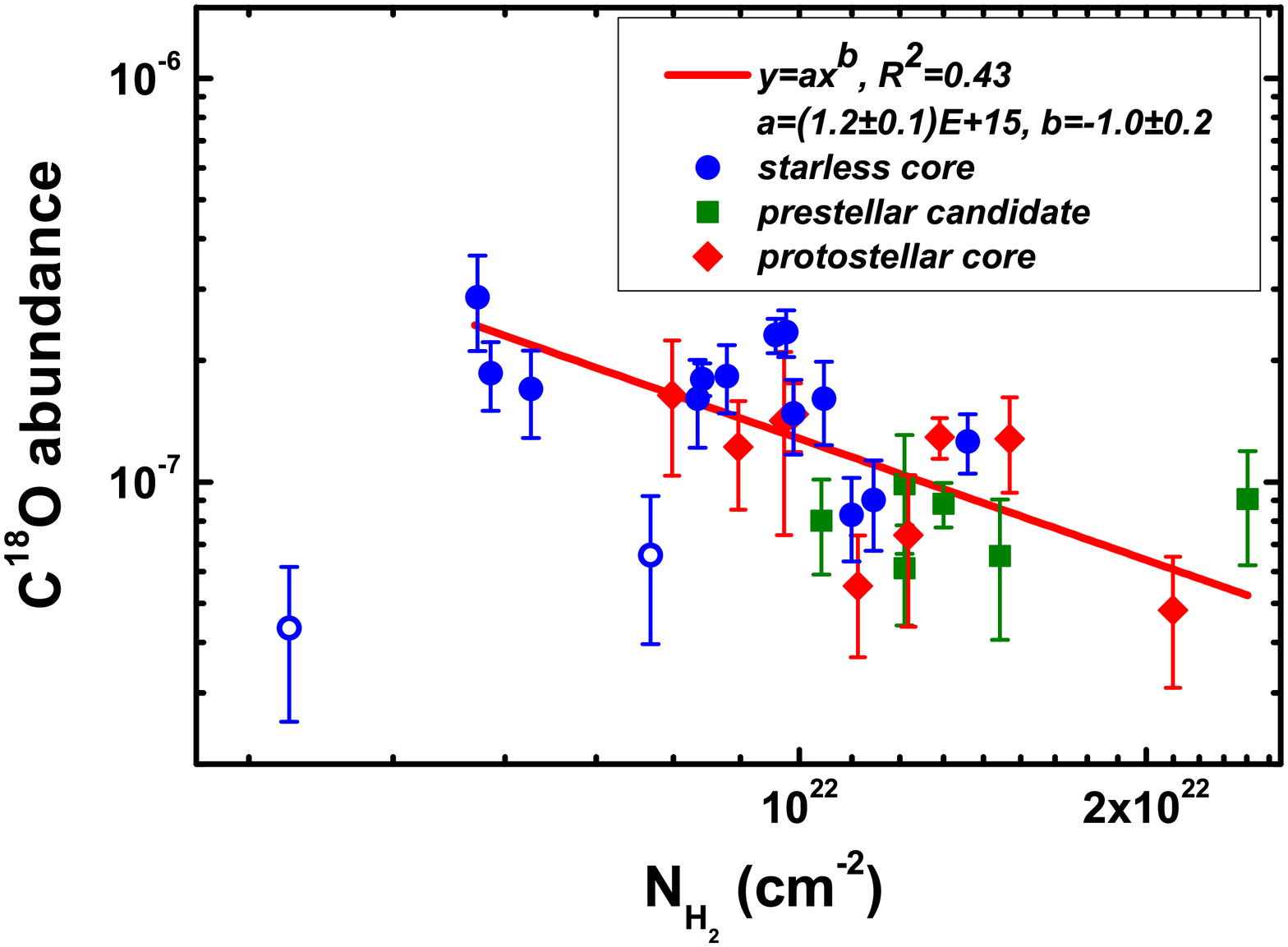}
\caption{Correlation of C$^{18}$O abundance and \emph{Herschel} $N_{\rm H_{2}}$. The red line is a power-law fit. The red diamonds, green squares and blue dots represent the protostellar cores, prestellar core candidates and starless coress, respectively. The blue circles represent G166.99-15.34-H1 and G167.23-15.32-H1, they are located outside of the densest parts of the filament, Thus, we excluded them from fitting.
The function and coefficients of models are presented in upper-right corner of panel. A potential anti-correlation are found in the figure.}
\label{}
\end{figure}
\begin{figure}[!ht] 
\plotone{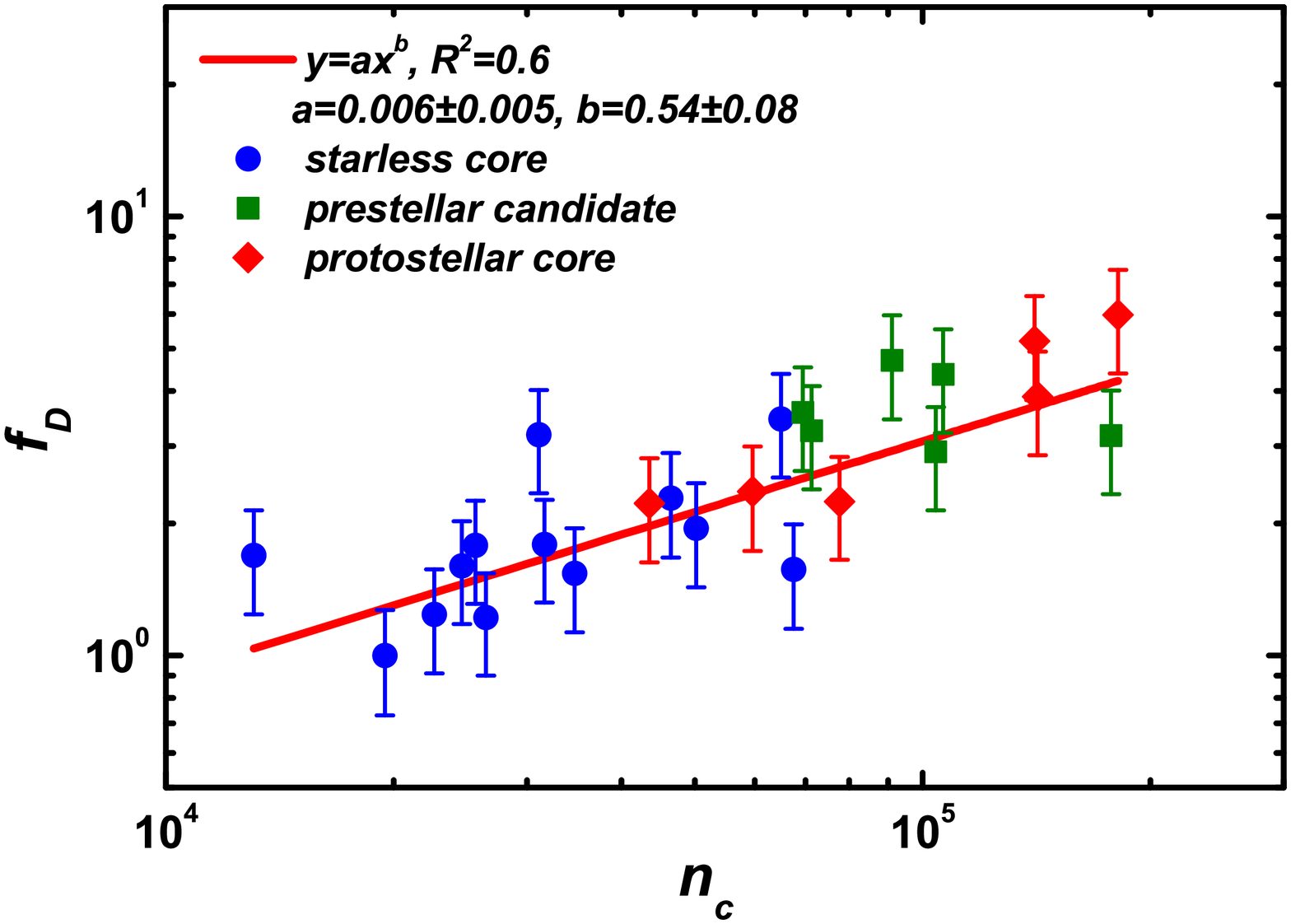}
\caption{Correlation of relative depletion factors ($f_{\rm D}$) and the central volume densities $n_{\rm c}$ (\emph{Herschel}).
The red line is a power-law fit, blue dots, green squares and red diamonds represent starless cores, prestellar core candidates and protostellar cores, respectively.
}
\label{}
\end{figure}

\section{SUMMARY}
To understand better the properties of PGCCs and dense cores in L1495 cloud, we have studied 16 dense clumps in the L1495 cloud with data from \emph{Herschel}, JCMT/SCUBA-2, and the PMO 13.7 m telescope. The main findings of this work are as follows:
\begin{enumerate}
\item In the L1495 region of Taurus, we identified 30 dense cores in 16 PGCCs. The majority of Planck clumps have Herschel cores, often multiple, but only a subset of the Herschel cores have corresponding SCUBA-2 condensations. Based on SCUBA-2 and \emph{Herschel} 70 $\micron$ data, we have classified these 30 dense cores into three types: 15 are starless cores (i.e., with neither SCUBA-2 nor \emph{Herschel} 70 $\micron$ emission), 6 are prestellar core candidates (i.e., with SCUBA-2 emission but without \emph{Herschel} 70 $\micron$ emission) and 9 are protostellar cores (i.e., with \emph{Herschel} 70 $\micron$ emission). Our findings suggest that not all PGCCs contain prestellar objects. In general, however, the dense cores in PGCCs are usually at their earliest evolutionary stages.
\item Based on \emph{Herschel} data, the mean volume densities of the starless cores, prestellar core candidates and protostellar cores are 1.0($\pm$0.4)$\times10^{4}$ cm$^{-3}$ cm$^{-3}$, 2.5($\pm$0.6)$\times10^{4}$ cm$^{-3}$, and 2.7($\pm$1.0)$\times10^{4}$, respectively. The mean dust temperatures of the starless cores, prestellar core candidates and protostellar cores are 12.1($\pm$0.4) K, 11.6($\pm$0.8) K, and 12.4($\pm$0.7) K, respectively. On average, prestellar core candidates have slightly lower dust temperature and higher density than starless cores.
    The 9 protostellar cores appear to be still at the earliest protostellar evolutionary phases and the newly formed protostar have not significantly heated their envelopes.
\item Non-thermal and thermal velocity dispersions have been derived from the PMO data. In all 30 dense cores,
    the Mach numbers are all larger than 1, with an average value of 2.1($\pm$0.8). Hence the turbulence in the L1495 dense cores is supersonic, and all 30 dense cores may be turbulence-dominated.
\item The virial masses ($M_{\rm vir}$) derived from the C$^{18}$O data range from 3.6($\pm$0.7) M$_{\sun}$ to 74.3($\pm$7.6) M$_{\sun}$.
    The virial parameters ($\alpha$) range from 0.3($\pm$0.1) to 17.5($\pm$7.3).
     Fourteen dense cores have virial parameters smaller than 2, and 18 dense cores have virial parameters larger than 2. The virial parameter and core mass follow the power law trend, $\alpha$ = $\alpha_{\rm min}\cdot$($M_{\rm core}$/$M_{\rm max}$)$^{h_{\alpha}}$, with $h_{\alpha}$ = -0.74($\pm$0.08).
\item The column density profiles of 28 dense cores have been derived from \emph{Herschel} data. The central volume density ($n_{\rm c}$), the size of flat region (\emph{a}) and the truncated radius ($R_{\rm t}$)
      were obtained by fitting a column density profile. We found that the values of flat region size and truncated radius decrease as $n_{\rm c}$ increases. This indicates that dense cores shrink as they evolve from starless cores to protostellar cores \citep{Ward-Thompson94}.
\item The C$^{18}$O abundances are used to investigate the CO depletion degree in the three types of dense cores.
      The mean C$^{18}$O abundances of protostellar cores, prestellar core candidates, and starless cores are 1.12($\pm$0.48)$\times10^{-7}$, 8.07($\pm$1.34)$\times10^{-8}$ and 1.57($\pm$0.65)$\times10^{-7}$, respectively.
      This variation means that CO depletion in prestellar core candidates is most significant. The core-averaged CO depletion factors ($f_{\rm D}$) range from 1 to 6.6($\pm$1.8).
      Our results support the idea that the C$^{18}$O abundance can be used as an evolutionary tracer for molecular cloud cores, as suggested by \citet{Caselli99,Di Francesco07,Liu13}.
\end{enumerate}

\section{ACKNOWLEDGEMENT}
S.-L. Qin is supported by the National Key R\&D Program of China (NO. 2017YFA0402701), by the Joint Research Fund in Astronomy (U1631237) under cooperative agreement between the National Natural Science Foundation of China (NSFC) and Chinese Academy of Sciences (CAS), by the Top Talents Program of Yunnan Province (2015HA030).
Tie Liu is supported by KASI fellowship and EACOA fellowship.
Ke Wang is supported by grant WA3628-1/1 of the German Research Foundation (DFG) through the priority program 1573 (``Physics of the Interstellar Medium''). Chang Won Lee was supported by Basic Science Research Program through the National Research Foundation of Korea (NRF) funded by the Ministry of Education, Science and Technology (NRF-2016R1A2B4012593)
This research was partly supported by the OTKA grant NN-111016.
We thank Michel Fich, Mika Juvela, Jan Wouterloot, Archana Soam, M. R. Cunningham, Chang Won Lee, Paul F. Goldsmith, A. Rivera-Ingraham, Jinghua Yuan, Pak Shing Li, Johanna Malinen, George J. Bendo, Hong-Li Liu, Miju Kang, Neal J. Evans II, Patricio Sanhueza, Edith Falgarone, Glenn J. White, Izaskun Jimenez-Serra, You-Hua Chu, Yao-Lun Yang, JinHua He, and Hauyu Baobab Liu for helpful discussions and comments.
The James Clerk Maxwell Telescope is operated by the East Asian Observatory on behalf of The National Astronomical Observatory of Japan, Academia Sinica Institute of Astronomy and Astrophysics, the Korea Astronomy and Space Science Institute, the National Astronomical Observatories of China and the Chinese Academy of Sciences (Grant No. XDB09000000), with additional funding support from the Science and Technology Facilities Council of the United Kingdom and participating universities in the United Kingdom and Canada. Additional funds for the construction of SCUBA-2 were provided by the Canada Foundation for Innovation. This research has made use of data from the Herschel Gould Belt survey (HGBS) project (http://gouldbelt-herschel.cea.fr). The HGBS is a Herschel Key Programme jointly carried out by SPIRE Specialist Astronomy Group 3 (SAG 3), scientists of several institutes in the PACS Consortium (CEA Saclay, INAF-IFSI Rome and INAF-Arcetri, KU Leuven, MPIA Heidelberg), and scientists of the Herschel Science Center (HSC).

\clearpage



\end{document}